\documentclass{ptephy_v1}

\preprintnumber{KUNS-2833,YITP-20-103}

\usepackage{bm,amssymb, color}
\usepackage{amsmath, amsthm, mathrsfs}
\usepackage{braket}
\usepackage{cases}
\usepackage{txfonts}
\newcommand{\e}{e}
\newcommand{\ii}{i}

\begin{document}
\title{Microscopic derivation of density functional theory 
for superfluid systems based on effective action formalism}

\author{Takeru Yokota}
\affil{Institute for Solid State Physics, The University of Tokyo, Kashiwa, Chiba 277-8581, Japan}
\author{Haruki Kasuya}
\affil[2]{Yukawa Institute for Theoretical Physics, Kyoto University, Kyoto 606-8502, Japan}
\author{Kenichi Yoshida}
\affil[3]{Department of Physics, Kyoto University, Kyoto 606-8502, Japan}
\author[2]{Teiji Kunihiro}


\begin{abstract}
Density-functional theory for superfluid systems is developed 
in the framework of the functional renormalization group based on the effective action formalism.
We introduce the effective action for
the particle-number and nonlocal pairing densities
and demonstrate that the Hohenberg--Kohn theorem for superfluid systems
is established in terms of the effective action.
The flow equation for the effective action is then derived, 
where the flow parameter runs from $0$ to $1$, corresponding to the non-interacting and interacting systems. 
From the flow equation and the variational equation 
that the equilibrium density satisfies,
we obtain the exact 
expression for
the Kohn--Sham potential generalized to including the pairing potentials.
The resultant Kohn--Sham potential has a nice feature that it expresses the 
microscopic formulae 
of the external, Hartree, pairing, and exchange-correlation terms, separately.
It is shown that our 
Kohn--Sham potential gives 
the ground-state energy of the Hartree--Fock--Bogoliubov 
theory by neglecting the correlations. 
An advantage of our exact formalism lies in the fact 
that it provides ways to systematically improve the correlation part.
\end{abstract}

\subjectindex{A63, B32}

\maketitle

\section{Introduction}
The purpose of the present paper is to
develop a density-functional theory (DFT) for superfluid systems
based on the effective action formalism, where 
the Hohenberg--Kohn theorem is readily established for superfluid systems with 
the spontaneous symmetry breaking (SSB) being incorporated. 
Moreover, it will be shown that the exact expression for the Kohn--Sham (KS)
potential 
is derived by making use of the functional renormalization group (FRG).

DFT has been widely applied and 
proven to be an efficient and convenient framework to deal with the many-body problems 
in various fields including quantum chemistry and atomic, molecular, condensed-matter, and nuclear physics; 
see Refs.~\cite{coh12, lau13, mar17, jon15, dru09, nak16} for some recent reviews. 
DFT rests upon the Hohenberg--Kohn  theorem~\cite{hoh64} which states that 
the ground state of a quantum many-body system 
is obtained by minimizing the energy-density functional (EDF) with respect 
to the particle density only. 
In a practical implementation of DFT, one usually relies on the 
KS scheme~\cite{koh65}, where 
the ground-state density of the interacting system is reproduced by just solving 
the Hartree--Fock-type single-particle Schr\"odinger equation, 
called the 
KS equation, for the non-interacting reference system. 

However it is known to be clumsy for 
 a naive application of the original DFT to describe the systems with SSB 
such as magnetization or superfluidity solely with the particle density.
So  the order parameter as given by an anomalous density 
is included explicitly as an additional density to describe such a system with 
SSB \cite{bar72,pan72,oli88,lud05,mar05,lin15a,lin15b,bri05,bro13,dob84,jin17,kas20}.

A great challenge in the current study of DFT 
is to develop a microscopic and systematic framework for constructing the EDF 
from an inter-particle interaction.
The notion of the effective action~\cite{JonaLasinio:1964cw,Weinberg:1996kr} has brought forth 
the nonperturbative framework for dealing with the quantum many-body problems,
and the concepts and recipes developed and accumulated in the study of quantum-field theory (QFT) can 
be brought into the study of many-body systems.
In particular, the effective action for composite fields \cite{lut60,bay62,cor74} links DFT and concepts and recipes of QFT.
Indeed the two-particle point-irreducible 
effective action~\cite{ver92} for the density field 
is identified with the free-energy density functional 
containing all the information of not only the ground state but also the excited states~\cite{fuk94,val97a}.
Thus a promising strategy to overcome the perennial challenge has come out.
In Ref.~\cite{fuk88}, what they call the inversion method was developed to
calculate the ground-state energy nonperturbatively, and it was further extended 
to the Fermion system with superfluidity~\cite{ina92}.
Along with this line of method, the
notion of effective field theory is applied 
to construct the free-energy density functional including superfluidity~\cite{pug03,bha05,fur07,fur12a}.
 
Although the inversion method is certainly 
a well-founded and powerful nonperturbative scheme, 
the notions and techniques developed in QFT 
may not have been fully utilized.
An established method 
is currently available for the calculation of effective actions, which is 
called the functional renormalization group (FRG)~\cite{weg73,wil74,pol84,wet93}.
The FRG provides a nonperturbative and systematic procedure for
the analyses of renormalization flows 
by solving one-parameter functional differential equations 
in a closed form of effective actions.
Recently, the application to the 
effective action 
for composite fields aimed at an {\it ab-initio} construction of DFT,
which we call the functional-renormalization-group 
aided density functional theory (FRG-DFT), has been developed:
After the proposal of the formalism in Refs.~\cite{pol02,sch04},
several approximation schemes have been proposed,
whose performances were tested in (0+0)- or (0+1)-dimensional toy models~\cite{kem13,lia18}.
The application was extended to
(1+1)-dimensional systems composed of nucleons
in the canonical formalism~\cite{kem17a} and the grand-canonical formalism~\cite{yok18}.
And the excited states in (1+1)-dimensional systems were investigated~\cite{yok18b}.
Furthermore, analyses of more realistic systems such as
homogeneous electron gases in (2+1) dimensions~\cite{yok19}
and (3+1) dimensions~\cite{yok20} were recently achieved.

The purpose of this paper is to extend the FRG-DFT formalism as developed in Refs.~\cite{yok18,yok18b} 
so as to be applicable to superfluid systems at finite temperature.
To this end, we introduce the 
effective action for
the particle-number and nonlocal pairing densities, whereby
the Hohenberg--Kohn theorem 
in terms of the Helmholtz free energy is found to be readily established.
We then derive the flow equation for the effective action
by introducing a flow parameter $\lambda$ with 
which the strength of the inter-particle interaction is adiabatically increased
 as $\lambda$ is changed from $0$ to $1$.
The differentiation of  the effective action
with respect to $\lambda$ gives
the coupled flow equations for the particle-number and pairing densities.
Together with the variational equation which the equilibrium densities should satisfy,
the exact 
expression for the
KS potential generalized to including the pairing potentials is derived.
The resultant KS potential has a nice feature that it expresses the 
microscopic formulae 
of the external, Hartree, pairing, and exchange-correlation terms, separately.
As a demonstration of the validity of our formalism, we shall show that
a lowest-order approximation to our flow equations with respect to the flow parameter 
reproduces the gap equation, the ground-state energy and pairing density 
given in the BCS theory for
 the case of short-range interactions in the weak-coupling limit.
Although we show the validity of our microscopic formulation by applying it to simple cases,
an advantage of our exact FRG-DFT formalism lies in the fact  that it provides ways to systematically improve the correlation part.

This paper is organized as follows:
In Sec.~\ref{sec:formalism},
we introduce the effective action
for the particle-number and pairing densities
and derive the flow equation to calculate the effective action.
From the flow equation, the exact expression of
the 
KS potential is derived.
In Sec.~\ref{sec:lo}, we apply the lowest-order
approximation to our formalism and compare our results with those given by the BCS theory.
Section \ref{sec:conclusion} is devoted to a brief summary and the conclusions.

\section{Formalism \label{sec:formalism}}

In this section, we develop a formalism to describe microscopically the superfluid systems
at finite temperature in the framework of FRG-DFT, 
where the effective action given in the functional-integral form is fully utilized.
The formalism starts with construction of the 
effective action 
$\Gamma[\rho,\kappa,\kappa^*]$
for the  nonlocal pairing densities as well as particle-number ones,
and then we show that $\Gamma[\rho,\kappa,\kappa^*]$ can be
identified with the free-energy {\em density functional} 
for which the Hohenberg--Kohn theorem holds.
We then introduce a flow parameter $\lambda$ with 
which the inter-particle interaction is switched on gradually as $\lambda$ is varied from $0$ to $1$.
Differentiating thus constructed effective action $\Gamma_{\lambda}[\rho,\kappa,\kappa^*]$ 
with respect to $\lambda$, we obtain
the coupled flow equations for the particle-number and pairing densities. 
Thus the problem of deriving the governing equation of the 
superfluid systems in the context of DFT is reduced to solving the flow equations.
Inherently in the way of the introduction of the flow parameter,
the governing equation thus obtained actually leads to 
the exact expression of the 
KS potential.
Moreover, because of the very exact nature of the governing equation thus obtained,
the resultant equation can be a starting point for an introduction of various systematic
approximations.

We consider a system composed of
non-relativistic fermions interacting via two-body interactions 
subject to the external potential at finite temperature $1/\beta$
with $\beta$ being the inverse temperature.
The action describing the system in the imaginary-time formalism is given as follows:
\begin{align}
	\label{eq:action}
    S[\psi,\psi^*]
    =
    \int_{\xi}
    \psi^*
    (\xi+\epsilon_\tau)
    \left(
    \partial_\tau
    -
    \frac{\Delta}{2}
    +
    \mathcal{V}({\xi})
    \right)
    \psi(\xi)
    +
    \frac{1}{2}
    \int_{\xi,\xi'}
    \mathcal{U}(\xi,\xi')
    \psi^*(\xi+\epsilon_\tau)
    \psi^*(\xi'+\epsilon_\tau)
    \psi(\xi')
    \psi(\xi).
\end{align}
Here, the short-hand notations
$\xi=(\tau,{\bf x}, a)$,\,
$\xi'=(\tau',{\bf x}', a'),\,\xi'+\epsilon_\tau=(\tau'+\epsilon,{\bf x}', a')$
 and $\int_\xi=\sum_a\int_0^\beta d\tau \int d{\bf x}$ have been introduced,
where $\tau$, ${\bf x}$, and $a$ represent the imaginary time, spatial coordinate, 
and index for internal degree of freedoms, respectively with $\epsilon$ being 
an infinitesimal constant.
The fermion fields are represented by Grassmann variables $\psi(\xi)$ and $\psi^*(\xi)$.
The static external potential and the instantaneous two-body interaction
are denoted by $\mathcal{V}(\xi)=V_a({\bf x})$ and
$    \mathcal{U}(\xi,\xi') =  \delta(\tau - \tau')U_{aa'}({\bf x}-{\bf x}')$,
respectively. 
The shifted coordinate $\xi+\epsilon_\tau=(\tau+\epsilon, {\bf x},a)$ 
is introduced so that the corresponding Hamiltonian becomes normal-ordered \cite{yok18}.

\subsection{Effective action for superfluid systems \label{sec:ea_edf}}

Let us introduce
the effective action, which is  to be identified with the free-energy density 
functional for superfluid systems.
The construction of the effective action~\cite{JonaLasinio:1964cw,Weinberg:1996kr} 
starts by  defining the generating functional of the
correlation functions of density fields.
In this work, we generalize the previous works 
\cite{fuk94,yok18,yok18b} and introduce the generating functional for
the correlation functions of particle-number and (nonlocal) pairing densities as
\begin{align}
	\label{eq:wdef}
	Z[\vec{J}]
	=
	&
    \int \mathcal{D}\psi \mathcal{D}\psi^*
    e^{-S[\psi, \psi^*]
    +\int_\xi J^\rho(\xi)
    \psi^*(\xi+\epsilon_{\tau})\psi(\xi)
    +
    \int_{\xi,\xi'}J^\kappa(\xi,\xi')\psi(\xi)\psi(\xi')
    +
    \int_{\xi,\xi'}J^{\kappa^*}(\xi,\xi')\psi^*(\xi')\psi^*(\xi)
    }
    \notag
    \\
    =&
    \int \mathcal{D}\psi \mathcal{D}\psi^*
    e^{-S[\psi, \psi^*]
    +
    \sum_{i=\rho,\kappa,\kappa^*}
    \int_{\overline{\xi}^i} J^i(\overline{\xi}^i)
    \hat{\rho}_i(\overline{\xi}^i)}.
\end{align}
$\vec{J}=(J^{\rho}(\overline{\xi}^\rho),J^{\kappa}(\overline{\xi}^\kappa),J^{\kappa^*}(\overline{\xi}^{\kappa^*}))$
denotes the external source coupled to the respective density fields:
\begin{align}
	\label{eq:rhat}
    \left\lbrace
    \hat{\rho}_{i}(\overline{\xi}^i)
    \right\rbrace_{i=\rho,\kappa,\kappa^*}
    =
    \lbrace
    \psi^*(\xi+\epsilon_\tau)\psi(\xi)
    ,
    \psi(\xi)\psi(\xi')
    ,
    \psi^*(\xi')\psi^*(\xi)
    \rbrace.
\end{align}
with the following abbreviated notations for arguments
\begin{align}
	\overline{\xi}^i \equiv
	\left\{
	\begin{array}{ccccc}
	\xi &=&(\tau, {\bf x}, a) \quad \quad & ;&i=\rho \\
	(\xi, \xi')&=&(\tau, {\bf x}, a, \tau', {\bf x}', a') & ;&i=\kappa^{(*)} 
	\end{array}
	\right.
	.
\end{align}
We emphasize that the pairing fields are introduced as nonlocal ones with which
the present formalism is applicable for
describing not only the conventional spin-singlet pairing but any types of pairings 
as in Ref.~\cite{cap97}
and furthermore the UV divergence, which appears in the case of local interactions 
and local pairings, can be avoided.

Now we define the generating functional for connected correlation functions
$W[\vec{J}]$ by
\begin{align}
\label{eq:def-W}
{W[\vec{J}]} := \ln Z[\vec{J}].
\end{align}
Then the effective action 
$\Gamma[\vec{\rho}=(\rho,\kappa,\kappa^{*})]$
for the particle-number density $\rho(\xi)$ as well as nonlocal pairing densities
$\kappa^{(*)}(\xi,\xi')$
is given by a Legendre transformation of
$W[\vec{J}]$:
\begin{align}
    \label{eq:gamma}
	\Gamma[\vec{\rho}]
	&=
    \sup_{\vec{J}}
	\left(
	\sum_i \int_{\overline{\xi}^i}
	J^{i}(\overline{\xi}^i)\rho_i(\overline{\xi}^i)
	-
	W[\vec{J}]
	\right)
\notag	\\
	&=: \sum_i \int_{\overline{\xi}^i} 	
	J_{{\rm sup}}^i[\vec{\rho}](\overline{\xi}^i)
	\rho_i(\overline{\xi}^i)
	-
	W[\vec{J}_{\rm sup}[\vec{\rho}]].
\end{align}
Here $\vec{J}_{\rm sup}[\vec{\rho}]$ denotes the source that gives the supremum of the quantity 
in the bracket of the first line 
and gives the expectation value of the density operator $\hat{\rho}^i$
\begin{align}
	\label{eq:w1}
	\frac{\delta W[\vec{J}]}{\delta J^{i}(\overline{\xi}^i)}\Bigg\vert_
{\vec{J}=\vec{J}_{{\rm sup}}[\vec{\rho}]}
	=
	\rho_{i}(\overline{\xi}^i)
\end{align}
on account of Eqs.~\eqref{eq:wdef} and \eqref{eq:def-W}.

\subsection{The Hohenberg--Kohn theorem for superfluid systems}

Now that having defined the effective action by Eq.~\eqref{eq:gamma} for superfluid systems 
in a rather natural way, we are in a position to show 
one of the most important points in this article, i.e., a proof of the Hohenberg--Kohn theorem for superfluid systems with nonzero temperature \cite{oli88} on the basis of the effective action without recourse to any 
heuristic arguments.
\begin{itemize}
	\item {\bf Variational principle for Eq.~\eqref{eq:gamma}.}

	For a proper description of
    the superfluidity where the SSB is accompanied,
	it is most convenient to introduce a $\tau$-independent artificial external potential
	$\mu^{{\kappa^{(*)}}}(\overline{\xi}^{\kappa^{(*)}})$ 
	coupled to the pairing density $\kappa^{(*)}(\overline{\xi}^{\kappa^{(*)}})$, which will be, however, taken 
	to be zero in the end of calculation.
	We shall consider a generic case where
           the chemical potential $\mu_a$ may
        depend on the internal degree of freedom $a$; however, notice that 
    the possible spatial dependence of $\mu_a$ can be 
        neglected without loss of generality because it 
        can be absorbed away into the redefinition of $V_a({\bf x})$.

	Now under the condition that the external potential 
	$\delta_{i\rho}\mathcal{V}(\overline{\xi}^\rho)-\mu^i(\overline{\xi}^{i})$ is  applied to the system,
	consider the variational problem of the following functional:
	\begin{align}
		\label{eq:Ivari}
		I[\vec{\rho}]
		=
		\Gamma[\vec{\rho}]
		-
		\sum_{i}\int_{\overline{\xi}^i}\mu^i(\overline{\xi}^i)\rho_i(\overline{\xi}^i),
	\end{align}
	where $\vec{\mu}
	=(\mu^\rho(\overline{\xi}^\rho),
	\mu^{\kappa}(\overline{\xi}^{\kappa}),
	\mu^{\kappa^{*}}(\overline{\xi}^{\kappa^{*}})
	)
	=(\mu_a, 
	\mu^{\kappa}(\overline{\xi}^{\kappa}),
	\mu^{\kappa^{*}}(\overline{\xi}^{\kappa^{*}}))$.
        
	The solution $\vec{\rho}_{\rm ave}$ of the variational equation 
        $\delta I[\vec{\rho}]/ \delta \rho_i(\overline{\xi}^i)=0$
	satisfies $J_{\rm sup}^i[\vec{\rho}_{\rm ave}](\overline{\xi}^i)=\mu^i(\overline{\xi}^i)$,
        and thus one obtains
	\begin{align}
		\label{eq:rho_vari}
		\rho_{{\rm ave},i}(\overline{\xi}^i)
		=
		\frac{1}{Z[\vec{\mu}]}
		\int \mathcal{D}\psi \mathcal{D}\psi^*
		\hat{\rho}_i(\overline{\xi}^i)
		e^{-S[\psi, \psi^*]
		+
		\sum_{j=\rho,\kappa,\kappa^*}
		\int_{\overline{\xi}^{j}_1} \mu^{j}(\overline{\xi}^{j}_1)
		\hat{\rho}_{j}(\overline{\xi}^{j}_1)},
	\end{align}
	by the use of Eqs.~\eqref{eq:wdef} and \eqref{eq:w1}.
	Equation \eqref{eq:rho_vari} shows that
	$\vec{\rho}_{\rm ave}$ is the average of the density in the equilibrium state in the presence of the external potential $\delta_{i\rho}\mathcal{V}(\overline{\xi}^\rho)-\mu^i(\overline{\xi}^{i})$.
	
	A remark is in order here: 
	Although the vanishing limit is formally taken for  
	$\mu^{{\kappa^{(*)}}}(\overline{\xi}^{\kappa^{(*)}})$,
	this limit should be taken with a care when SSB is concerned;
        it should be interpreted as 
        $0+$ in the mathematical notation meaning
	an infinitesimally small but nonzero value to obtain $\vec{\rho}_{\rm ave}$ variationally.
	Figure \ref{fig:Ishape} shows a schematic picture of 
	$I[\vec{\rho}]$ solely as a function of $\kappa$, say, with $\rho$ fixed.
	If we have two solutions $\kappa_{\rm ave}^{(1)}$ and $\kappa_{\rm ave}^{(2)}$ as
        the pairing density  in equilibrium, 
	$I[\vec{\rho}]$ with $\mu^{{\kappa^{(*)}}}=0$ is actually
           flat in the interpolated densities 
	[$\kappa=\alpha \kappa_{\rm ave}^{(1)} + (1-\alpha)\kappa_{\rm ave}^{(2)}$, $0\leq \alpha \leq 1$]
	since $\Gamma[\vec{\rho}]$ is a convex functional as a result of the Legendre transformation.
	This prevents one from determining 
         the equilibrium density uniquely by a variational method.
	As long as
	$\mu^{{\kappa^{(*)}}}$ is kept small but nonvanishing, however,
	the flat region of $I[\vec{\rho}]$ is  slightly tilted, 
        then one can uniquely identify the equilibrium density 
        in the variational method.

	\begin{figure}[!t]
	    \centering
    	\includegraphics[width=0.9\columnwidth]{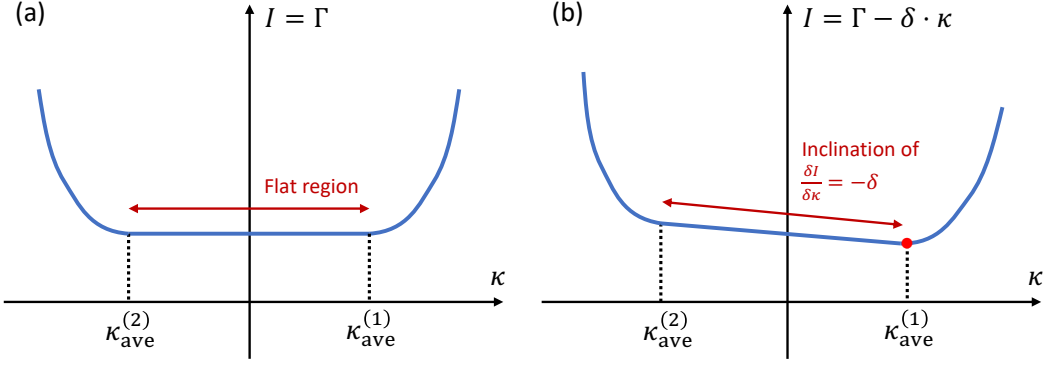}
    	\caption{Schematic pictures of the $\kappa$-dependence of $I$ 
    	in the case of (a) no external potential and 
    	(b) small nonzero external potential, where the $\rho$-dependence is not depicted.
    	Two different equilibrium pairing densities are denoted by $\kappa_{\rm ave}^{(1,2)}$.
    	$\delta$ is an infinitesimally small but nonzero number.
    	Without the external potential, $\kappa_{\rm ave}$ is not determined uniquely in the variational way since $I$ has the same value in the region shown in (a).
    	With a small external field being applied, the equilibrium density [$\kappa^{(1)}_{\rm ave}$ in (b)]
    	may be obtained variationally since the flat region is slightly tilted.
    	\label{fig:Ishape}}
    \end{figure}

    Furthermore, it follows immediately from Eq.~(\ref{eq:rho_vari}) that 
    the external potential determines  the equilibrium density uniquely.
    On the other hand, it is easily demonstrated that the equilibrium density 
    in turn determines 
    uniquely the external potential
    by observing that $I[\vec{\rho}_{\rm ave}]/\beta$ coincides with
    the grand potential  $\Omega[\vec{\mu}]=-W[\vec{\mu}]/\beta$
    for $\vec{\rho}_{\rm ave}$ minimizing $I[\vec{\rho}]$:  
    \begin{align}
	I[\vec{\rho}_{\rm ave}]=-W[\vec{J}_{\rm sup}[\vec{\rho}_{\rm ave}]]
	+\sum_i \int_{\overline{\xi}^i} [J_{\rm sup}^i [\vec{\rho}_{\rm ave}]
	(\overline{\xi}^i)-\mu^i(\overline{\xi}^i)]\rho_i(\overline{\xi}^i)
	=
	-W[\vec{\mu}],
     \end{align}
    where we have used 
    $J_{\rm sup}^i[\vec{\rho}_{\rm ave}](\overline{\xi}^i)=\mu^i(\overline{\xi}^i)$.
     This argument is based on the fact that the functional $I[\vec{\rho}]$
     is strictly convex as seen in Fig.~\ref{fig:Ishape}(b). 
     Therefore, one has a one-to-one mapping between the external potential 
     $\delta_{i\rho}\mathcal{V}(\overline{\xi}^\rho)
	-\mu^i(\overline{\xi}^{i})$ and the equilibrium density $\vec{\rho}_{\rm ave}$.
     In particular, the existence of the map
     from $\vec{\rho}_{\rm ave}$ to 
     $\delta_{i\rho}\mathcal{V}(\overline{\xi}^\rho)-\mu^i(\overline{\xi}^{i})$
     can be seen more clearly through the fact that $\Gamma[\vec{\rho}]$ 
     is linearly dependent on the external field, which we now show below.
     
	\item {\bf One-to-one map between
        the external fields and the densities.}
         
	From Eq.~\eqref{eq:gamma}, we obtain
	\begin{align}
		\label{eq:gdecomp}
		\Gamma[\vec{\rho}]
		=&
		\sup_{\vec{J}}
		\left(
		\sum_i \int_{\overline{\xi}^i}
		J^{i}(\overline{\xi}^i)\rho_i(\overline{\xi}^i)
		-
		\left.W\right|_{\mathcal{V}=0}
		[\vec{J}-(\mathcal{V},0,0)]
		\right)
		\notag
		\\
		=&
		\int_{\xi}\mathcal{V}({\xi})\rho({\xi})
		+
		\sup_{\vec{J}}
		\left(
		\sum_i \int_{\overline{\xi}^i}
		J^{i}(\overline{\xi}^i)\rho_i(\overline{\xi}^i)
		-
		\left.W\right|_{\mathcal{V}=0}[\vec{J}]
		\right)
		\notag
		\\
		=&
		\int_{\xi}\mathcal{V}({\xi})\rho({\xi})
		+
		\left.\Gamma\right|_{\mathcal{V}=0}[\vec{\rho}],
	\end{align}
	where $\left.W\right|_{\mathcal{V}=0}[\vec{J}]$ and $\left.\Gamma\right|_{\mathcal{V}=0}[\vec{\rho}]$
	are defined by the action with vanishing external potential.
	From the first to second line of Eq.~\eqref{eq:gdecomp}, we have shifted $\vec{J}$ 
          as $\vec{J}\to \vec{J}+(\mathcal{V},0,0)$.
	This shows that $\Gamma[\vec{\rho}]$ can be decomposed into the universal part
	$\left.\Gamma\right|_{\mathcal{V}=0}[\vec{\rho}]$ independent of $\mathcal{V}({\xi})$
	and the term depending on $\mathcal{V}({\xi})$ linearly.
	
	Through Eq.~\eqref{eq:gdecomp} and the variational equation of $I[\vec{\rho}]$,
	the existence of the one-to-one map from 
	$\vec{\rho}_{\rm ave}$
	to $\delta_{i\rho}\mathcal{V}(\overline{\xi}^\rho)
	-\mu^i(\overline{\xi}^{i})$ is established.
	To show this, let us assume that two sets of
	the external potential and the chemical potential denoted by
	$(\mathcal{V}_1, \vec{\mu}_1)$
	and
	$(\mathcal{V}_2, \vec{\mu}_2)$, respectively,
	give the same densities $\vec{\rho}_{\rm ave}$.
	By use of the variational equation of $I[\vec{\rho}]$ and Eq.~\eqref{eq:gdecomp}, 
        however, we obtain
	\begin{align*}
		\frac{\delta \Gamma|_{\mathcal{V}=0}[\vec{\rho}_{\rm ave}]}{\delta \rho_i(\overline{\xi}^i)}
		+
		\delta_{i\rho}\mathcal{V}_1(\overline{\xi}^\rho)
		-
		\mu_1^i(\overline{\xi}^i)
		=0,
		\\
		\frac{\delta \Gamma|_{\mathcal{V}=0}[\vec{\rho}_{\rm ave}]}{\delta \rho_i(\overline{\xi}^i)}
		+
		\delta_{i\rho}\mathcal{V}_2(\overline{\xi}^\rho)
		-
		\mu_2^i(\overline{\xi}^i)
		=0.
	\end{align*}
	Subtracting these equations from each other,
	we have the equality
	\begin{align}
		\delta_{i\rho}\mathcal{V}_1(\overline{\xi}^\rho)
		-
		\mu_1^i(\overline{\xi}^i)
		=
		\delta_{i\rho}\mathcal{V}_2(\overline{\xi}^\rho)
		-
		\mu_2^i(\overline{\xi}^i),
	\end{align}
	which proves that $\delta_{i\rho}\mathcal{V}(\overline{\xi}^\rho)-\mu^i(\overline{\xi}^i)$ 
        is uniquely determined by $\vec{\rho}_{\rm ave}$.
\end{itemize}

Thus the Hohenberg--Kohn theorem for superfluid systems with nonzero temperature, 
which was first demonstrated in Ref.~\cite{oli88}, 
is established in terms of the effective action.
Furthermore it is also evident that
$\beta^{-1}\Gamma[\vec{\rho}]$ can be identified with Helmholtz free energy 
$F_{\rm H}[\vec{\rho}]$  at $\vec{\rho}=\vec{\rho}_{\rm ave}$ as
\begin{align}
	\label{eq:f_gam}
	F_{\rm H}[\vec{\rho}]
	=
	\frac{\Gamma[\vec{\rho}]}{\beta}.
\end{align}
In fact, in the limit of $\mu^{\kappa^{(*)}} \to 0$, 
Eq.~\eqref{eq:f_gam} at $\vec{\rho}=\vec{\rho}_{\rm ave}$ becomes
\begin{align}
	\label{eq:f_omega}
	F_{\rm H}[\vec{\rho}_{\rm ave}]
	=
	\frac{1}{\beta}
	\int_{\xi}\mu_a\rho_{\rm ave}(\xi)
    -
    \frac{1}{\beta}
	W[\vec{\mu}=(\mu_a, 0, 0)],
\end{align}
because of the relation $\vec{J}_{\rm sup}[\vec{\rho}_{\rm ave}]=\vec{\mu}$ and Eq.~\eqref{eq:gamma}.
In terms of words,
since $Z[\vec{\mu}]$ is the grand partition function in the presence of the chemical potential $\vec{\mu}$, 
$-W[\vec{\mu}]/\beta=-\ln Z[\mu]/\beta$ is the grand potential, and hence
 $F_{\rm H}[\vec{\rho}_{\rm ave}]$ in Eq.~\eqref{eq:f_omega} is identified with the Helmholtz free energy.

Finally, let us consider the case of zero-temperature limit.
At the zero temperature limit $\beta\to \infty$,
$F_{\rm H}[\vec{\rho}_{{\rm ave}}]$ becomes the ground-state energy $E_{{\rm gs}}$ 
because $F_{\rm H}[\vec{\rho}_{{\rm ave}}]$ can be written as 
$F_{\rm H}[\vec{\rho}_{{\rm ave}}]=-\beta^{-1}\ln \sum_n \exp(-\beta E_{n})$,
where $\lbrace E_{n} \rbrace$ is the energy eigenvalues of the system
and satisfies $E_{\rm gs}=E_{0}<E_{1}<\cdots$.
Therefore, we can identify the EDF with the effective action as
\begin{align}
	E[\vec{\rho}]=\lim_{\beta\to \infty} \frac{1}{\beta} \Gamma[\vec{\rho}].
\end{align}
In particular, the universal part of the EDF is obtained by substitution 
of $\Gamma$ with $\Gamma|_{\mathcal{V}=0}$.
This concludes the generalization of the correspondence between the EDF and the effective action 
 to the superfluid systems~\cite{fuk94,val97a,yok18}.

\subsection{Flow equation} \label{sec:flow}

The effective action $\Gamma[\vec{\rho}]$ contains the whole contents of physics of the quantum system 
in a compact form of the functional integral. Although the compact and exact form of the action allows us to 
define consistent approximations,
the problem is developing a computational method to perform  the functional integral  in a systematic and
 hopefully exact way. In this subsection,
we are going to develop a practical method
 to calculate the effective action in such a manner.

A wise way originally developed by 
Wegner \cite{weg73} and Wilson \cite{wil74} 
is to convert the functional integral to a differential equation, which is called a renormalization-group (RG)
equation or flow equation, and solve the equation with some systematic approximation 
being adopted.
In the case of the effective action solely for the particle-number density
and accordingly without SSB, a calculational framework is developed 
based on  an RG/flow equation in Refs.~\cite{pol02,sch04}; see also 
Refs.~\cite{yok18,yok18b}.
Here, we generalize this framework to the case with SSB
for the calculation of $\Gamma[\vec{\rho}]$
containing the pairing density fields.

Following Refs.~\cite{pol02,sch04,yok18,yok18b}, 
we introduce a regulated two-body interaction $\mathcal{U}_\lambda(\xi, \xi')$
with a flow parameter $\lambda$ and define  the regulated action as
\begin{align}
	\label{eq:reg_S}
    S_\lambda[\psi,\psi^*]
    =
    \int_{\xi}
    \psi^*
    (\xi+\epsilon_\tau)
    \left(
    \partial_\tau
    -
    \frac{\Delta}{2}
    +
    \mathcal{V}(\xi)
    \right)
    \psi(\xi)
    +
    \frac{1}{2}
    \int_{\xi,\xi'}
    \mathcal{U}_{\lambda}(\xi,\xi')
    \psi^*(\xi+\epsilon_\tau)
    \psi^*(\xi'+\epsilon_\tau)
    \psi(\xi')
    \psi(\xi),
\end{align}
where $\mathcal{U}_\lambda(\xi, \xi')$ is defined by
\begin{align}
\label{eq:def-of-cal-U-lambda}
	\mathcal{U}_\lambda(\xi, \xi') = 
	\delta(\tau - \tau') U_{\lambda, aa'}({\bf x} - {\bf x}'),
\end{align}
with a constraint
\begin{align*}	
	U_{\lambda=0,aa'}({\bf x} - {\bf x}') = &
	0,
	\\
	U_{\lambda=1,aa'}({\bf x} - {\bf x}') = &
	U_{aa'}({\bf x} - {\bf x}')
\end{align*}
so that the action evolves from the free $S_{\lambda=0}$
to the fully-interacting one $S_{\lambda=1}$ as $\lambda$ runs from 0 to 1.
With this action, the $\lambda$-dependent
generating functionals and effective action
are defined in the same manner as in Sec.~\ref{sec:ea_edf}:
\begin{align}
	\label{eq:w_lam}
	W_\lambda[\vec{J}]
	=&
	\ln Z_\lambda[\vec{J}]
    =
    \ln \int \mathcal{D}\psi \mathcal{D}\psi^*
    e^{-S_\lambda[\psi, \psi^*]
    +
    \sum_{i=\rho,\kappa,\kappa^*}
    \int_{\overline{\xi}^i} J^i(\overline{\xi}^i)
    \hat{\rho}_i(\overline{\xi}^i)},
	\\
	\label{eq:gam_lam}
	\Gamma_\lambda[\vec{\rho}]
	=&
    \sup_{\vec{J}}
	\left(
	\sum_i \int_{\overline{\xi}^i}
	J^{i}(\overline{\xi}^i)\rho_i(\overline{\xi}^i)
	-
	W_\lambda[\vec{J}]
	\right)
	=:
	\sum_i \int_{\overline{\xi}^i} 	
	J_{{\rm sup},\lambda}^i[\vec{\rho}](\overline{\xi}^i)
	\rho_i(\overline{\xi}^i)
	-
	W_\lambda[\vec{J}_{{\rm sup},\lambda}[\vec{\rho}]],
\end{align}
where $J_{{\rm sup},\lambda}^i[\vec{\rho}](\overline{\xi}^i)$ satisfies
\begin{align}
	\label{eq:w1l}
	\frac{\delta W_\lambda[\vec{J}]}{\delta J^{i}(\overline{\xi}^i)}\Bigg\vert_
{\vec{J}=\vec{J}_{{\rm sup,\lambda}}[\vec{\rho}](\overline{\xi})}
	=
	\rho_{i}(\overline{\xi}^i).
\end{align}


Now utilizing the $\lambda$ dependence of the action (\ref{eq:gam_lam}),
 the flow equation of $\Gamma_\lambda[\vec{\rho}]$
can be obtained simply by differentiating it with respect to $\lambda$ \cite{pol02,sch04,kem17a,yok18,yok18b}.
It turns out, however that
the most convenient way to obtain the flow equation for $\Gamma_\lambda[\vec{\rho}]$
is first to derive that for $W_{\lambda}[\vec{J}]$ 
instead of $\Gamma_\lambda[\vec{\rho}]$.
The differentiation of $W_{\lambda}[\vec{J}]$ with respect to $\lambda$ reads
\begin{align}
	\label{eq:dw1}
	\partial_\lambda W_{\lambda}[\vec{J}]
	=&
	-
    \frac{1}{2}
    \int_{\overline{\xi}^\rho_1:=(\tau_1,{\bf x}_1 ,a_1),
    \overline{\xi}^\rho_2:=(\tau_2,{\bf x}_2 ,a_2)}
    \partial_\lambda
    \mathcal{U}_{\lambda}(\overline{\xi}^\rho_1,\overline{\xi}^\rho_2)
    \langle
    \psi^*(\overline{\xi}^\rho_1+\epsilon_\tau)
    \psi^*(\overline{\xi}^\rho_2+\epsilon_\tau)
    \psi(\overline{\xi}^\rho_2)
    \psi(\overline{\xi}^\rho_1)
    \rangle_{\lambda,\vec{J}},
\end{align}
where
\begin{align}
	\label{eq:ave_def}
	\langle
	\mathcal{O}[\psi,\psi^*]
	\rangle_{\lambda,\vec{J}}
	:=
	\frac{1}{Z_{\lambda}[\vec{J}]}
    \int \mathcal{D}\psi \mathcal{D}\psi^*
    \mathcal{O}[\psi,\psi^*]
    e^{-S_{\lambda}[\psi, \psi^*]
    +
    \sum_{i=\rho,\kappa,\kappa^*}
    \int_{\overline{\xi}^i} \vec{J}(\overline{\xi}^i)
    \hat{\rho}_i(\overline{\xi}^i)},
\end{align}
with $\mathcal{O}[\psi,\psi^*]$ being some functional of $\psi$ and $\psi^*$.
Noting that $\mathcal{U}_{\lambda}(\overline{\xi}^\rho_1,\overline{\xi}^\rho_2)$ defined in 
Eq.~\eqref{eq:def-of-cal-U-lambda} involves the delta function of times, the integrand in Eq.~\eqref{eq:dw1}
can be cast into the form
of the density correlation functions with the use of the  commutation relation:
\begin{align}
	&
    \partial_\lambda
    \mathcal{U}_{\lambda}(\overline{\xi}^\rho_1,\overline{\xi}^\rho_2)
    \langle
    \psi^*(\overline{\xi}^\rho_1+\epsilon_\tau)
    \psi^*(\overline{\xi}^\rho_2+\epsilon_\tau)
    \psi(\overline{\xi}^\rho_2)
    \psi(\overline{\xi}^\rho_1)
    \rangle_{\lambda,\vec{J}}
    \notag
    \\
     = &
    \delta(\tau_1-\tau_2) \partial_\lambda U_{\lambda, aa'}({\bf x} - {\bf x}')
	\langle
    \psi^*(\overline{\xi}^\rho_1+\epsilon_\tau)
    \psi^*(\overline{\xi}^\rho_2+\epsilon_\tau)
    \psi(\overline{\xi}^\rho_2)
    \psi(\overline{\xi}^\rho_1)
    \rangle_{\lambda,\vec{J}}
    \notag
    \\
    = &
    \delta(\tau_1-\tau_2) \partial_\lambda U_{\lambda, aa'}({\bf x} - {\bf x}')
    \Bigl[
	\langle
    \psi^*(\overline{\xi}^\rho_1+\epsilon_\tau'+\epsilon_\tau)
    \psi(\overline{\xi}^\rho_1+\epsilon_\tau')
    \psi^*(\overline{\xi}^\rho_2+\epsilon_\tau)
    \psi(\overline{\xi}^\rho_2)
    \rangle_{\lambda,\vec{J}} 
    \notag      \\
   \quad    & \hspace{4cm} - 
    \delta_{a_1,a_2}
	\langle
    \psi^*(\overline{\xi}^\rho_1+\epsilon_\tau)
    \psi(\overline{\xi}^\rho_1)
    \rangle_{\lambda,\vec{J}}
    \delta({\bf x}_1-{\bf x}_2)
    \Bigr],
\end{align}
where 
an infinitesimal number $\epsilon_\tau'$
plays the same role as $\epsilon_\tau$
but its zero limit should be taken after that of $\epsilon_{\tau}$; 
see Appendix in Ref.~\cite{yok18}.
From Eq.~\eqref{eq:w_lam}, one can immediately see
\begin{align}
	\label{eq:ravew1}
	\langle
    \hat{\rho}_{i}(\overline{\xi}^i)
    \rangle_{\lambda,\vec{J}}
    =&
	\frac{\delta W_{\lambda}[\vec{J}]}{\delta J_i(\overline{\xi}^i)},
    \\
    \label{eq:fourpt_w}
	\langle
    \psi^*(\overline{\xi}^\rho_1+\epsilon_\tau'+\epsilon_\tau)
    \psi(\overline{\xi}^\rho_1+\epsilon_\tau')
    \psi^*(\overline{\xi}^\rho_2+\epsilon_\tau)
    \psi(\overline{\xi}^\rho_2)
    \rangle_{\lambda,\vec{J}}
    =&
	\frac{\delta W_{\lambda}[\vec{J}]}{\delta J_\rho(\overline{\xi}^\rho_1)}
	\frac{\delta W_{\lambda}[\vec{J}]}{\delta J_\rho(\overline{\xi}^\rho_2)}
	+
	\frac{\delta^2 W_{\lambda}[\vec{J}]}
	{\delta J_\rho(\overline{\xi}^\rho_1+\epsilon_\tau')
	\delta J_\rho(\overline{\xi}^\rho_2)}.
\end{align}
Using these relations, one finds that Eq.~\eqref{eq:dw1}
takes the form of the flow equation for $W_{\lambda}[\vec{J}]$ as follows:
\begin{align}
	\label{eq:wflow}
	\partial_\lambda W_{\lambda}[\vec{J}]
	=&
	-
    \frac{1}{2}
    \int_{\overline{\xi}^\rho_1,\overline{\xi}^\rho_2}
    \partial_\lambda
    \mathcal{U}_{\lambda}(\overline{\xi}^\rho_1,\overline{\xi}^\rho_2)
    \notag
    \\
    &\times
    \left(
	\frac{\delta W_{\lambda}[\vec{J}]}{\delta J_\rho(\overline{\xi}^\rho_1)}
	\frac{\delta W_{\lambda}[\vec{J}]}{\delta J_\rho(\overline{\xi}^\rho_2)}
	+
	\frac{\delta^2 W_{\lambda}[\vec{J}]}
	{\delta J_\rho(\overline{\xi}^\rho_1+\epsilon_\tau')
	\delta J_\rho(\overline{\xi}^\rho_2)}    
    -
	\frac{\delta W_{\lambda}[\vec{J}]}{\delta J_\rho(\overline{\xi}^\rho_1)}
    \delta_{a_1,a_2}\delta({\bf x}_1-{\bf x}_2)
    \right).
\end{align}

Next we shall show that this flow equation is in turn  converted to that for $\Gamma_\lambda[\vec{\rho}]$
on account of Eq.~\eqref{eq:gam_lam}.
By differentiating Eq.~\eqref{eq:w1l} with respect to $\vec{J}_{\rm sup, \lambda}[\vec{\rho}]$, we have
\begin{align}
	\label{eq:w2}
	\frac{\delta^2 W_\lambda[\vec{J}_{\rm sup, \lambda}[\vec{\rho}]]}
	{\delta J_{i}(\overline{\xi}^i_1)\delta J_{j}(\overline{\xi}^j_2)}
	=
	\left.
	\frac{\delta \rho_{i}(\overline{\xi}^i_1)}{\delta J_{j}(\overline{\xi}^j_2)}
	\right|_{\vec{J}=\vec{J}_{\rm sup, \lambda}[\vec{\rho}]}.
\end{align}
By taking the second derivative of 
Eq.~\eqref{eq:gamma} with respect to $\vec{\rho}$, we have
\begin{align}
	\label{eq:gam2}
	\frac{\delta^2 \Gamma_\lambda[\vec{\rho}]}
	{\delta \rho_{i}(\overline{\xi}^i_1)\delta \rho_{j}(\overline{\xi}^j_2)}
	=&
	\frac{\delta J_{{\rm sup}, \lambda}^j[\vec{\rho}](\overline{\xi}^j_2)}{\delta \rho_{i}(\overline{\xi}^i_1)}.
\end{align}
One finds that Eq.~\eqref{eq:w2} is the inverse of Eq.~\eqref{eq:gam2}:
\begin{align}
	\label{eq:w2g2}
	\frac{\delta^2 W_\lambda[\vec{J}_{\rm sup, \lambda}[\vec{\rho}]]}
	{\delta J_{i}(\overline{\xi}^i_1)\delta J_{j}(\overline{\xi}^j_2)}
	=
	\left(
	\frac{\delta^2 \Gamma_\lambda}{\delta \vec{\rho} \delta \vec{\rho}}
	\right)^{-1}_{ij}
	[\vec{\rho}]
	(\overline{\xi}^i_1,\overline{\xi}^j_2).
\end{align}
Here, $(\delta^2\Gamma_\lambda/\delta\vec{\rho}\delta\vec{\rho})^{-1}_{ij}[\vec{\rho}](\overline{\xi}^i,\overline{\xi'}^j)$
is defined as follows:
\begin{align}
	\label{eq:gam2_inv}
    \sum_{k=\rho,\kappa,\kappa^{*}}
    \int_{\overline{\xi''}^k}
    \left(
    \frac{\delta^2\Gamma_\lambda}{\delta\vec{\rho}\delta\vec{\rho}}
    \right)^{-1}_{ik}[\vec{\rho}](\overline{\xi}^i,\overline{\xi''}^k)
    \frac{\delta^2\Gamma_\lambda [\vec{\rho}]}
    {\delta \rho_k(\overline{\xi''}^k) \delta\rho_j(\overline{\xi'}^j)}
    =
    \delta(\overline{\xi}^i,\overline{\xi'}^j),
\end{align}
and $\delta(\overline{\xi}^i,\overline{\xi'}^j)$ is defined 
by
\begin{align*}
	\delta(\overline{\xi}^i,\overline{\xi'}^j)
	=
	\begin{cases}
	\delta(\xi,\xi')
	&
	(i=j=\rho,\,\overline{\xi}^\rho=\xi,\,\overline{\xi}^\rho=\xi')
	\\
	\delta(\xi_1,\xi'_1)\delta(\xi_2,\xi'_2)
	&
	(i=j=\kappa^{(*)},\,\overline{\xi}^{\kappa^{(*)}}=(\xi_1,\xi_2),\,
	\overline{\xi'}^{\kappa^{(*)}}=(\xi_1',\xi_2'))
	\\
	0
	&
	({\text{otherwise}})
	\end{cases},
\end{align*}
with $\delta(\xi=(\tau,{\bf x},a),\xi'=(\tau',{\bf x}',a'))=\delta_{a,a'}\delta(\tau-\tau')\delta({\bf x}-{\bf x}')$.
By taking the derivative of 
Eq.~\eqref{eq:gamma} with respect to $\lambda$, we obtain  
\begin{align}
	\label{eq:dgdw}
	\partial_{\lambda}\Gamma_\lambda[\vec{\rho}]
	=
	-\left(\partial_\lambda W_\lambda\right)[\vec{J}_{\rm sup, \lambda}[\vec{\rho}]].
\end{align}
Substituting Eqs.~\eqref{eq:w1l}, \eqref{eq:w2g2}, and \eqref{eq:dgdw}
into Eq.~\eqref{eq:wflow}, we eventually arrive at 
the formal flow equation for $\Gamma_\lambda[\vec{\rho}]$:
\begin{align}
	\label{eq:gflow}
	\partial_\lambda \Gamma_\lambda[\vec{\rho}]
	=&
	\frac{1}{2}\int_{\overline{\xi}^\rho_1=(\tau_1,{\bf x}_1,a_1),
	\overline{\xi}^\rho_2=(\tau_2,{\bf x}_2,a_2)}
	\partial_\lambda\mathcal{U}_{\lambda}(\overline{\xi}^\rho_1,\overline{\xi}^\rho_2)
	\notag
	\\
	&\times\left[
	\rho(\overline{\xi}^\rho_1)\rho(\overline{\xi}^\rho_2)
	+
	\left(
	\frac{\delta^2 \Gamma_\lambda}{\delta \vec{\rho} \delta \vec{\rho}}
	\right)^{-1}_{\rho\rho}
	[\vec{\rho}]
	(\overline{\xi}^\rho_1+{\epsilon}_{\tau}',\overline{\xi}^\rho_2)
	-
	\rho(\overline{\xi}^\rho_1)
	\delta_{a_1,a_2}
	\delta({\bf x}_1-{\bf x}_2)
	\right].
\end{align}

Now 
Eqs.~\eqref{eq:fourpt_w} and \eqref{eq:w2g2} together with
Eq.~\eqref{eq:w1l} tell us that 
$(\frac{\delta^2 \Gamma_\lambda}{\delta \vec{\rho} \delta \vec{\rho}})^{-1}_{\rho\rho}$
 gives essentially the correlation functions of the densities $\vec{\rho}$, 
which collectively denote the particle-number and pairing densities.
To see more 
explicitly the physical contents of the flow equation (\ref{eq:gflow}),
in particular,
 the contributions of the pairing densities, we find it 
convenient to separate
the $\lambda=0$ part of the correlation function where the particle operators are free ones:
\begin{align*}
	&\left(
	\frac{\delta^2 \Gamma_\lambda}{\delta \vec{\rho} \delta \vec{\rho}}
	\right)^{-1}_{\rho\rho}
	[\vec{\rho}]
	(\overline{\xi}^\rho_1+{\epsilon}_{\tau}',\overline{\xi}^\rho_2)
	-
	\rho(\overline{\xi}^\rho_1)
	\delta_{a_1,a_2}
	\delta({\bf x}_1-{\bf x}_2)
	\\
	=
	&
	\langle
	\psi^*(\overline{\xi}^\rho_1+\epsilon_\tau+\epsilon'_\tau)
	\psi(\overline{\xi}^\rho_1+\epsilon'_\tau)
	\psi^*(\overline{\xi}^\rho_2+\epsilon_\tau)
	\psi(\overline{\xi}^\rho_2)
	\rangle_{\lambda,\vec{J}_{\rm sup, \lambda}[\vec{\rho}]}
	-
	\rho(\overline{\xi}^\rho_1+\epsilon'_\tau)
	\rho(\overline{\xi}^\rho_2)
	-
	\rho(\overline{\xi}^\rho_1)
	\delta_{a_1,a_2}
	\delta({\bf x}_1-{\bf x}_2)
	\\
	=
	&
	\langle
	\psi^*(\overline{\xi}^\rho_1+\epsilon_\tau+\epsilon'_\tau)
	\psi(\overline{\xi}^\rho_1+\epsilon'_\tau)
	\psi^*(\overline{\xi}^\rho_2+\epsilon_\tau)
	\psi(\overline{\xi}^\rho_2)
	\rangle_{\lambda=0,\vec{J}_{\rm sup, \lambda=0}[\vec{\rho}]}
	-
	\rho(\overline{\xi}^\rho_1+\epsilon'_\tau)
	\rho(\overline{\xi}^\rho_2)
	-
	\rho(\overline{\xi}^\rho_1)
	\delta_{a_1,a_2}
	\delta({\bf x}_1-{\bf x}_2)
	\\
	&+\mathcal{O}(\lambda).
\end{align*}
By 
applying  the Wick's theorem, 
we 
obtain
\begin{align}
	\label{eq:gamma_decomp}
	&\left(
	\frac{\delta^2 \Gamma_\lambda}{\delta \vec{\rho} \delta \vec{\rho}}
	\right)^{-1}_{\rho\rho}
	[\vec{\rho}]
	(\overline{\xi}^\rho_1+{\epsilon}_{\tau}',\overline{\xi}^\rho_2)
	-
	\rho(\overline{\xi}^\rho_1)
	\delta_{a_1,a_2}
	\delta({\bf x}_1-{\bf x}_2)
	\notag
	\\
	=
	&
	\kappa^*(\overline{\xi}_{12}^{\kappa^*})
	\kappa(\overline{\xi}_{12}^{\kappa})
	|_{\overline{\xi}_{12}^{\kappa^{(*)}}=({\overline{\xi}^\rho_1,\overline{\xi}^\rho_2})}
	\notag
	\\
	&-
	\langle
	\psi^*(\overline{\xi}^\rho_1+\epsilon_\tau+\epsilon'_\tau)
	\psi(\overline{\xi}^\rho_2)
	\rangle_{\lambda=0,\vec{J}_{\rm sup, \lambda=0}[\vec{\rho}]}
	\langle
	\psi^*(\overline{\xi}^\rho_2+\epsilon_\tau)
	\psi(\overline{\xi}^\rho_1+\epsilon'_\tau)
	\rangle_{\lambda=0,\vec{J}_{\rm sup, \lambda=0}[\vec{\rho}]}
	-
	\rho(\overline{\xi}^\rho_1)
	\delta_{a_1,a_2}
	\delta({\bf x}_1-{\bf x}_2)
	\notag
	\\
	&+\mathcal{O}(\lambda),
\end{align}
where we have used the equality 
$\langle\hat{\rho}_{i}(\overline{\xi}^i)\rangle_{\lambda,\vec{J}_{\rm sup,\lambda}[\vec{\rho}]}=\rho_{i}(\overline{\xi}^i)$
which follows from Eqs.~\eqref{eq:ravew1} and \eqref{eq:w1l}.
One sees that the first term 
given by a product of the pairing condensates in 
Eq.~(\ref{eq:gamma_decomp}) is 
the counter part to that of the particle-number densities 
given in the first term of Eq.~(\ref{eq:gflow}),
while the terms in the second line 
 and the remaining order-$\lambda$ terms
are identified 
with the exchange and correlation terms, respectively.
Thus one finds that it is natural to  represent Eq.~\eqref{eq:gflow} in a form where
the terms with different physical significance are written separately as,
\begin{align}
	\label{eq:gflod}
	\partial_\lambda \Gamma_\lambda[\vec{\rho}]
	=&
	\frac{1}{2}\int_{\overline{\xi}^\rho_1=(\tau_1,{\bf x}_1,a_1),
	\overline{\xi}^\rho_2=(\tau_2,{\bf x}_2,a_2)}
	\partial_\lambda\mathcal{U}_{\lambda}(\overline{\xi}^\rho_1,\overline{\xi}^\rho_2)
	\notag
	\\
	&\times\left[
	\rho(\overline{\xi}^\rho_1)\rho(\overline{\xi}^\rho_2)
	+
	\kappa^*(\overline{\xi}_{12}^{\kappa^*})
	\kappa(\overline{\xi}_{12}^{\kappa})
	|_{\overline{\xi}_{12}^{\kappa^{(*)}}=({\overline{\xi}^\rho_1,\overline{\xi}^\rho_2})}
	+
	G^{(2)}_{{\rm xc},\lambda}[\vec{\rho}]({\overline{\xi}^\rho_1,\overline{\xi}^\rho_2})
	\right],
\end{align}
where 
$G^{(2)}_{{\rm xc,\lambda}}[\vec{\rho}]({\overline{\xi}^\rho_1,\overline{\xi}^\rho_2})$
denotes the exchange--correlation term
given by 
\begin{align}
	\label{eq:g2xc}
	G^{(2)}_{{\rm xc,\lambda}}[\vec{\rho}]({\overline{\xi}^\rho_1,\overline{\xi}^\rho_2})
	=
	G^{(2)}_{{\rm x}}[\vec{\rho}]({\overline{\xi}^\rho_1,\overline{\xi}^\rho_2})
	+
	G^{(2)}_{{\rm c,\lambda}}[\vec{\rho}]({\overline{\xi}^\rho_1,\overline{\xi}^\rho_2}),
\end{align}
with
\begin{align}
	\label{eq:g2x}
	G^{(2)}_{{\rm x}}[\vec{\rho}]({\overline{\xi}^\rho_1,\overline{\xi}^\rho_2})
	=&
	\left(
	\frac{\delta^2 \Gamma_{\lambda=0}}{\delta \vec{\rho} \delta \vec{\rho}}
	\right)^{-1}_{\rho\rho}
	[\vec{\rho}]
	(\overline{\xi}^\rho_1+{\epsilon}_{\tau}',\overline{\xi}^\rho_2)
	-
	\kappa^*(\overline{\xi}_{12}^{\kappa^*})
	\kappa(\overline{\xi}_{12}^{\kappa})
	|_{\overline{\xi}_{12}^{\kappa^{(*)}}=({\overline{\xi}^\rho_1,\overline{\xi}^\rho_2})}
	-
	\rho(\overline{\xi}^\rho_1)
	\delta_{a_1,a_2}
	\delta({\bf x}_1-{\bf x}_2),
	\\
	\label{eq:g2correlation}
	G^{(2)}_{{\rm c},\lambda}[\vec{\rho}]({\overline{\xi}^\rho_1,\overline{\xi}^\rho_2})
	=&
	\left(
	\frac{\delta^2 \Gamma_\lambda}{\delta \vec{\rho} \delta \vec{\rho}}
	\right)^{-1}_{\rho\rho}
	[\vec{\rho}]
	(\overline{\xi}^\rho_1+{\epsilon}_{\tau}',\overline{\xi}^\rho_2)
	-
	\left(
	\frac{\delta^2 \Gamma_{\lambda=0}}{\delta \vec{\rho} \delta \vec{\rho}}
	\right)^{-1}_{\rho\rho}
	[\vec{\rho}]
	(\overline{\xi}^\rho_1+{\epsilon}_{\tau}',\overline{\xi}^\rho_2).
\end{align}
We call
$G^{(2)}_{{\rm x}}[\vec{\rho}]({\overline{\xi}^\rho_1,\overline{\xi}^\rho_2})$
and
$G^{(2)}_{{\rm c},\lambda}[\vec{\rho}]({\overline{\xi}^\rho_1,\overline{\xi}^\rho_2})$
the exchange and correlation terms, respectively.

We note that the flow equation gives the systematic way to calculate
the higher-order contribution for $\Gamma_{\lambda}[\vec{\rho}]$, or $G^{(2)}_{\rm c, \lambda}[\vec{\rho}]$, 
since it is not only an exact but also closed equation for $\Gamma_\lambda[\vec{\rho}]$.
Moreover, it also provides us with the basis for defining consistent approximations, say,
respecting symmetries, if necessary.
In this respect, some approximation schemes developed for the study of the normal phase
such as the vertex expansion \cite{pol02, sch04} is expected to be utilized
for the determination of $\Gamma_\lambda[\vec{\rho}]$ also in the case of superfluid systems.

\subsection{Expressions of Helmholtz free energy and ground-state energy}

Having derived the flow equation (\ref{eq:gflod}) for 
the effective action $\Gamma_\lambda[\vec{\rho}]$,
we here give a reduced formula for the Helmholtz free energy which is given in terms of the effective action.

The Helmholtz free energy given in Eq.~\eqref{eq:f_gam} now takes the following form
\begin{align}
	\label{eq:fave_gam}
	F_{\rm H}[\vec{\rho}_{\rm ave}]=\frac{\Gamma_{\lambda=1}[\vec{\rho}_{\rm ave}]}{\beta},
\end{align}
where $\vec{\rho}_{\rm ave}$ is the equilibrium density of the fully-interacting ($\lambda=1$) system in the presence of a given chemical potential $\vec{\mu}$:
\begin{align}
	\label{eq:rho_ave}
	\rho_{{\rm ave},i}(\overline{\xi}^i)
	=
	\frac{1}{Z_{\lambda=1}[\vec{\mu}]}
	\int \mathcal{D}\psi \mathcal{D}\psi^*
	\hat{\rho}_i(\overline{\xi}^i)
	e^{-S_{\lambda=1}[\psi, \psi^*]
	+
	\sum_{j=\rho,\kappa,\kappa^*}
	\int_{\overline{\xi}^j_1} \mu^j(\overline{\xi}^j_1)
	\hat{\rho}_j(\overline{\xi}^j_1)},
\end{align}
which is nothing but Eq.~\eqref{eq:rho_vari} with a replacement $Z[\vec{\mu}]\,\to\,Z_{\lambda=1}[\vec{\mu}]$.
The expression for $\Gamma_{\lambda=1}[\vec{\rho}_{\rm ave}]$ is obtained
by integrating Eq.~\eqref{eq:gflod} with respect to $\lambda$. 
Inserting the expression into Eq.~\eqref{eq:fave_gam} and using Eq.~\eqref{eq:gdecomp}
for $\Gamma_{\lambda=0}$, we have
\begin{align}
	\label{eq:fave_exp}
	F_{\rm H}[\vec{\rho}_{\rm ave}]
	=&
	\frac{\left.\Gamma_{\lambda=0}\right|_{\mathcal{V}=0}[\vec{\rho}_{\rm ave}]}{\beta}
	+
	\frac{1}{\beta}\int_{\xi} \mathcal{V}(\xi)\rho_{\rm ave}(\xi)
	\notag
	\\
	&
	+
	\frac{1}{2\beta}
	\int_{\overline{\xi}^\rho_1,\overline{\xi}^\rho_2}
	\mathcal{U}_{\lambda=1}(\overline{\xi}^\rho_1,\overline{\xi}^\rho_2)
	\left[
	\rho_{\rm ave}(\overline{\xi}^\rho_1)\rho_{\rm ave}(\overline{\xi}^\rho_2)
	+
	\kappa^*_{\rm ave}(\overline{\xi}_{12}^{\kappa^*})
	\kappa_{\rm ave}(\overline{\xi}_{12}^{\kappa})
	|_{\overline{\xi}_{12}^{\kappa^{(*)}}=({\overline{\xi}^\rho_1,\overline{\xi}^\rho_2})}
	\right]
	\notag
	\\
	&+
	\int_0^1 d\lambda
	\frac{1}{2\beta}
	\int_{\overline{\xi}^\rho_1,\overline{\xi}^\rho_2}
	\partial_\lambda\mathcal{U}_{\lambda}(\overline{\xi}^\rho_1,\overline{\xi}^\rho_2)
	G^{(2)}_{{\rm xc},\lambda}[\vec{\rho}_{\rm ave}]({\overline{\xi}^\rho_1,\overline{\xi}^\rho_2}).
\end{align}

The ground-state energy is obtained by taking the zero temperature limit in Eq.~\eqref{eq:fave_exp}:
\begin{align}
	\label{eq:egs_exp}
	E_{\rm gs}
    =&\lim_{\beta\to \infty}F_{\rm H}[\vec{\rho}_{\rm ave}]
	\notag \\
	=&
	T[\vec{\rho}_{\rm gs}]
	+
	\lim_{\beta\to\infty}
	\frac{1}{\beta}\int_{\xi} \mathcal{V}(\xi)\rho_{\rm ave}(\xi)
	\notag
	\\
	&
	+
	\lim_{\beta\to \infty}
	\frac{1}{2\beta}
	\int_{\overline{\xi}^\rho_1,\overline{\xi}^\rho_2}
	\mathcal{U}_{\lambda=1}(\overline{\xi}^\rho_1,\overline{\xi}^\rho_2)
	\left[
	\rho_{\rm ave}(\overline{\xi}^\rho_1)\rho_{\rm ave}(\overline{\xi}^\rho_2)
	+
	\kappa^*_{\rm ave}(\overline{\xi}_{12}^{\kappa^*})
	\kappa_{\rm ave}(\overline{\xi}_{12}^{\kappa})
	|_{\overline{\xi}_{12}^{\kappa^{(*)}}=({\overline{\xi}^\rho_1,\overline{\xi}^\rho_2})}
	\right]
	\notag
	\\
	&+
	\lim_{\beta\to \infty}
	\int_0^1 d\lambda
	\frac{1}{2\beta}
	\int_{\overline{\xi}^\rho_1,\overline{\xi}^\rho_2}
	\partial_\lambda\mathcal{U}_{\lambda}(\overline{\xi}^\rho_1,\overline{\xi}^\rho_2)
	G^{(2)}_{{\rm xc},\lambda}[\vec{\rho}_{\rm ave}]({\overline{\xi}^\rho_1,\overline{\xi}^\rho_2}),
\end{align}
where $T[\vec{\rho}_{\rm gs}]
:=
\lim_{\beta\to\infty}\left.\Gamma_{\lambda=0}\right|_{\mathcal{V}=0}[\vec{\rho}_{\rm ave}]/\beta$
with $\vec{\rho}_{\rm gs} \coloneqq \lim_{\beta\to\infty} \vec{\rho}_{\rm ave}$ being the ground-state density
is interpreted as the kinetic energy. 
Notice that $T[\vec{\rho}_{\rm gs}]$ is neither the kinetic energy of a free gas nor 
that of the interacting system, but that of the non-interacting system with an appropriate potential, as may be utilized in 
KS formalism.
Indeed, we shall show in the following subsection that the 
KS theory for superfluid systems at finite temperature 
emerges naturally in the present formalism.

\subsection{The Kohn--Sham potential \label{sec:selfcon}}

In our formulation of the FRG-DFT theory for superfluid systems, the starting action
with $\lambda=0$ is that for a non-interacting system with the external forces ensuring
the densities. Thus it is natural that
$\vec{J}_{\rm sup,\lambda}[\vec{\rho}]$
is related to the 
KS potential $\vec{\mathcal{V}}_{\rm KS}[\vec{\rho}]$
that is generalized so as
to incorporate the anomalous or pair potentials 
besides the normal ones: The 
KS potential is defined by the derivative of the EDF
subtracted by the non-interacting kinetic energy part at zero
temperature. 
In terms of the effective action at finite temperature, 
the kinetic energy part $T_{\rm KS}[\vec{\rho}]$ 
may be identified with $\Gamma_{\lambda=0}|_{\mathcal{V}=0}$,
and then we naturally arrive at the definition of
the 
KS potential as
\begin{align}
	\label{eq:ks_def}
	\mathcal{V}^i_{\rm KS}[\vec{\rho}](\overline{\xi}^i)
	=
	\frac{\delta}{\delta \rho_i(\overline{\xi}^i)}
	\left(
	\Gamma_{\lambda=1}[\vec{\rho}]-
\Gamma_{\lambda=0}|_{\mathcal{V}=0}[\vec{\rho}]
	\right).
\end{align}
By differentiating Eq.~\eqref{eq:gam_lam} with respect to the densities, we obtain
\begin{align}
	\label{eq:g1l}
	\frac{\delta \Gamma_{\lambda} [\vec{\rho}]}{\delta \rho_i(\overline{\xi}^i)}
	=
	J_{\rm sup,\lambda}^i[\vec{\rho}](\overline{\xi}^i).
\end{align}
Using this relation and Eq.~\eqref{eq:gdecomp} for $\Gamma=\Gamma_{\lambda=0}$,
Eq.~\eqref{eq:ks_def} is rewritten as follows:
\begin{align}
	\label{eq:ks_j}
	\mathcal{V}^i_{\rm KS}[\vec{\rho}](\overline{\xi}^i)
	=
	\delta_{i\rho}\mathcal{V}(\overline{\xi}^i)
	+
	J_{\rm sup,\lambda=1}^{i}[\vec{\rho}]
	-
	J_{\rm sup,\lambda=0}^{i}[\vec{\rho}].
\end{align}
The 
KS potential determines $\vec{\rho}_{\rm ave}$ for given chemical potentials
$\vec{\mu}$: Since $J_{\rm sup,\lambda=1}^i[\vec{\rho}_{\rm ave}](\overline{\xi}^i)=\mu^i(\overline{\xi}^i)$
is satisfied, Eq.~\eqref{eq:ks_j} at $\vec{\rho}=\vec{\rho}_{\rm ave}$ reads
\begin{align}
	\label{eq:ks_vari}
	\frac{\delta \Gamma_{\lambda=0}|_{\mathcal{V}=0}[\vec{\rho}_{\rm ave}]}{\delta \rho_i(\overline{\xi}^i)}
	+
	\mathcal{V}^i_{\rm KS}[\vec{\rho}_{\rm ave}](\overline{\xi}^i)
	=
	\mu^i(\overline{\xi}^i),
\end{align}
which 
is equivalent to the variational equation in the 
KS DFT~\cite{eng11}
and determines $\vec{\rho}_{\rm ave}$ from the 
KS potential (\ref{eq:ks_def}).
The present formulation does not refer to any single-particle orbitals for establishing the 
KS DFT.
Note that Eq.~(\ref{eq:ks_vari}) does possess the property of a self-consistency in terms of the densities.
In fact the equilibrium density of the full-interacting system 
with a given chemical potential $\vec{\mu}$ is also
 the equilibrium density 
of the non-interacting system under the external potential 
$\vec{\mu}-\vec{\mathcal{V}}_{\rm KS}[\vec{\rho}_{\rm ave}]$, as we will now show explicitly: 
First of all, we notice that Eq.~\eqref{eq:ks_vari} is also expressed as
$
J^i_{\rm sup, \lambda=0}[\vec{\rho}_{\rm ave}](\overline{\xi}^i)
	=
	\delta_{i\rho}\mathcal{V}(\overline{\xi}^i)
	+
	\mu^i(\overline{\xi}^i)
	-
	\mathcal{V}^i_{\rm KS}[\vec{\rho}_{\rm ave}](\overline{\xi}^i)
$
on account of the relation $J_{\rm sup,\lambda=1}^i[\vec{\rho}_{\rm ave}](\overline{\xi}^i)=\mu^i(\overline{\xi}^i)$ and 
Eq.~(\ref{eq:ks_j}).
Then inserting this relation into
 Eqs.~\eqref{eq:w_lam} and \eqref{eq:w1l}, one finds that Eq.~\eqref{eq:w1l} tells us
 that the equilibrium density $\vec{\rho}_{\rm ave}$ in Eq.~\eqref{eq:rho_ave} also 
has the following expression
\begin{align}
	\label{eq:rhoave_ks}
	\rho_{{\rm ave},i}(\overline{\xi}^i)
	=
	\frac{1}{\left.Z_{\lambda=0}\right|_{\mathcal{V}=0}[\vec{\mu}-\vec{\mathcal{V}}_{\rm KS}[\vec{\rho}_{\rm ave}]]}
    \int \mathcal{D}\psi \mathcal{D}\psi^*
    \hat{\rho}_i(\overline{\xi}^i)
    e^{-\left.S_{\lambda=0}\right|_{\mathcal{V}=0}[\psi^*, \psi]
    +
    \sum_{j=\rho,\kappa,\kappa^*}
    \int_{\overline{\xi}^j_1} 
    \left(
    \mu^j(\overline{\xi}^j_1)-\mathcal{V}^j_{\rm KS}[\vec{\rho}_{\rm ave}](\overline{\xi}^j_1)
    \right)
    \hat{\rho}_j(\overline{\xi}^j_1)},
\end{align}
where $\left.Z_{\lambda}\right|_{\mathcal{V}=0}[\vec{J}]$ is defined by the action 
$\left.S_{\lambda}\right|_{\mathcal{V}=0}[\psi^*, \psi]$
with no external potential in Eq.~\eqref{eq:w_lam}.
Note that Eq.~\eqref{eq:rhoave_ks} is a self-consistent equation for $\vec{\rho}_{\rm ave}$
since $\vec{\mathcal{V}}_{\rm KS}[\vec{\rho}_{\rm ave}]$ 
in the right-hand side depends on $\vec{\rho}_{\rm ave}$ in the left-hand side, as noted before.

Next, let us reduce the exact expression of 
the central quantity $\vec{\mathcal{V}}_{\rm KS}[\vec{\rho}]$ 
in our formalism in a more calculable form
from the flow equation \eqref{eq:gflod}.
After differentiating Eq.~\eqref{eq:gflod} with respect to $\vec{\rho}$
and then integrating it with respect to $\lambda$, we have
\begin{align}
	\label{eq:jcond}
    {J}_{{\rm sup},\lambda=1}^i[\vec{\rho}](\overline{\xi}^{i})
    =
    &
    {J}_{{\rm sup},\lambda=0}^i[\vec{\rho}](\overline{\xi}^{i})
    +
    \delta_{i,\rho}
	\int_{\overline{\xi}^\rho_1}
	\mathcal{U}_{\lambda=1}(\overline{\xi}^\rho,\overline{\xi}^\rho_1)
	\rho(\overline{\xi}^\rho_1)
    +
    \frac{\delta_{i,\kappa}}{2}
	\overline{\mathcal{U}}_{\lambda=1}(\overline{\xi}^{\kappa^{*}})
	\kappa^*(\overline{\xi}^{\kappa^{*}})
    +
    \frac{\delta_{i,\kappa^*}}{2}
	\overline{\mathcal{U}}_{\lambda=1}(\overline{\xi}^{\kappa})
	\kappa(\overline{\xi}^{\kappa})
	\notag
	\\
	&
	+
	\frac{1}{2}
	\int_0^1 d\lambda
	\int_{\overline{\xi}^{\rho}_1,	\overline{\xi}^{\rho}_2}
	\partial_\lambda
	\mathcal{U}_{\lambda}(\overline{\xi}^\rho_1,\overline{\xi}^\rho_2)
    \frac{\delta 
    G_{\rm xc,\lambda}^{(2)}[\vec{\rho}]
    (\overline{\xi}^\rho_1,\overline{\xi}^\rho_2)}
    {\delta \rho_{i}(\overline{\xi}^{i})},
\end{align}
where $\overline{\mathcal{U}}_{\lambda}(\overline{\xi}^{\kappa^{(*)}}:=(\xi,\xi'))
=\mathcal{U}_{\lambda}(\xi,\xi')$ and Eq.~\eqref{eq:g1l} has been used. 
With use of Eq.~\eqref{eq:ks_j},
we finally arrive at the following expression for $\vec{\mathcal{V}}_{\rm KS}[\vec{\rho}]$:
\begin{align}
	\label{eq:ks_g2}
    \mathcal{V}_{\rm KS}^i[\vec{\rho}]
    (\overline{\xi}^i)
    =
    &
    \delta_{i\rho}
    \mathcal{V}(\overline{\xi}^\rho)
    +
    \delta_{i,\rho}
	\int_{\overline{\xi}^\rho_1}
	\mathcal{U}_{\lambda=1}(\overline{\xi}^\rho,\overline{\xi}^\rho_1)
	\rho(\overline{\xi}^\rho_1)
    +
\left(    \frac{\delta_{i,\kappa}}{2}
	\overline{\mathcal{U}}_{\lambda=1}(\overline{\xi}^{\kappa^{*}})
	\kappa^*(\overline{\xi}^{\kappa^{*}})
    +
    \frac{\delta_{i,\kappa^*}}{2}
	\overline{\mathcal{U}}_{\lambda=1}(\overline{\xi}^{\kappa})
	\kappa(\overline{\xi}^{\kappa})
\right) 
	\notag
	\\
	&
	+
	\frac{1}{2}
	\int_0^1 d\lambda
	\int_{\overline{\xi}^{\rho}_1,	\overline{\xi}^{\rho}_2}
	\partial_\lambda
	\mathcal{U}_{\lambda}(\overline{\xi}^\rho_1,\overline{\xi}^\rho_2)
    \frac{\delta 
    G_{\rm xc,\lambda}^{(2)}[\vec{\rho}]
    (\overline{\xi}^\rho_1,\overline{\xi}^\rho_2)}
    {\delta \rho_{i}(\overline{\xi}^{i})}.
\end{align}
It is noteworthy that
 $\mathcal{V}_{\rm KS}^i[\vec{\rho}]$ is now naturally decomposed into 
four terms, which denote
the external, Hartree, pairing, and exchange-correlation terms, respectively.
In particular, a microscopic expression of the exchange-correlation term $\mathcal{V}_{\rm KS, xc}^i[\vec{\rho}]$
is explicitly given in terms of $G_{\rm xc,\lambda}^{(2)}[\vec{\rho}]$:
\begin{align}
	\mathcal{V}_{\rm KS, xc}^i[\vec{\rho}](\overline{\xi}^i)
	=
	\frac{1}{2}
	\int_0^1 d\lambda
	\int_{\overline{\xi}^{\rho}_1,	\overline{\xi}^{\rho}_2}
	\partial_\lambda
	\mathcal{U}_{\lambda}(\overline{\xi}^\rho_1,\overline{\xi}^\rho_2)
    \frac{\delta 
    G_{\rm xc,\lambda}^{(2)}[\vec{\rho}]
    (\overline{\xi}^\rho_1,\overline{\xi}^\rho_2)}
    {\delta \rho_{i}(\overline{\xi}^{i})}.
\end{align}

We emphasize that our FRG-DFT formalism applied to superfluid systems 
has naturally led to the self-consistent equation for the densities that are kept 
the same between the non-interacting and interacting systems together with the microscopic expressions of
the 
KS potential and the kinetic energy without recourse to notions of single-particle orbitals.
It is also noted here that the effective action formalism with the inversion method 
was applied to a microscopic construction of the 
KS potential 
for a normal system~\cite{val97} and a superfluid system of the local singlet pairs~\cite{fur07}. 
However, the exact expression of the 
KS potential is not derived in the inversion method. 
Furthermore, the single-particle basis are introduced in developing the 
KS DFT. 
\footnote{In Ref.~\cite{fur12a}, it was anticipated that 
the 
KS DFT may be constructed
without single-particle orbitals in the FRG-DFT.}

In the next section, we are going to show that
Eq.~\eqref{eq:fave_exp} together with Eqs.~\eqref{eq:ks_vari} and \eqref{eq:ks_g2} leads to
  the ground-state energy in the Hartree--Fock--Bogoliubov approximation,
which is furthermore reduced to
 the well-known gap equations of BCS theory 
 in the case of short-range interaction.

\section{Analysis in the lowest order approximation: neglect of correlations \label{sec:lo}}

We demonstrate that our formalism actually reproduces the results of the 
mean field theory at the lowest-order truncation,
implying that $G_{\rm xc,\lambda}^{(2)}$
does not have $\lambda$ dependence: $G_{\rm xc,\lambda}^{(2)}\approx G_{\rm x}^{(2)}$.

\subsection{The case of a generic interaction}

Applying the lowest-order truncation to Eq.~\eqref{eq:ks_g2}, we have
\begin{align}
	\label{eq:jcond_lo}
	\mathcal{V}_{\rm KS}^i[\vec{\rho}](\overline{\xi}^i)
    =
    &
    \delta_{i,\rho}
    \left(
    \mathcal{V}(\overline{\xi}^\rho)
    +
	\int_{\overline{\xi}^\rho_1}
	\mathcal{U}_{\lambda=1}(\overline{\xi}^\rho,\overline{\xi}^\rho_1)
	\rho(\overline{\xi}^\rho_1)
	\right)
    +
    \frac{\delta_{i,\kappa}}{2}
	\overline{\mathcal{U}}_{\lambda=1}(\overline{\xi}^{\kappa})
	\kappa^*(\overline{\xi}^{\kappa})
    +
    \frac{\delta_{i,\kappa^*}}{2}
	\overline{\mathcal{U}}_{\lambda=1}(\overline{\xi}^{\kappa^*})
	\kappa(\overline{\xi}^{\kappa^{*}})
	\notag
	\\
	&
	+
	\frac{1}{2}
	\int_{\overline{\xi}^{\rho}_1,	\overline{\xi}^{\rho}_2}
	\partial_\lambda
	\mathcal{U}_{\lambda}(\overline{\xi}^\rho_1,\overline{\xi}^\rho_2)
    \frac{\delta 
    G_{\rm x}^{(2)}[\vec{\rho}]
    (\overline{\xi}^\rho_1,\overline{\xi}^\rho_2)}
    {\delta \rho_{i}(\overline{\xi}^{i})},
\end{align}
for the 
KS potential.
On account of Eq.~\eqref{eq:g2x},
${\delta G_{\rm x}^{(2)}[\vec{\rho}]
(\overline{\xi}^\rho_1,\overline{\xi}^\rho_2)}/{\delta \rho_{i_1}(\overline{\xi}^{i})}$
is represented as follows:
\begin{align}
	&\frac{\delta 
    G_{\rm x}^{(2)}
    [\vec{\rho}]
    (\overline{\xi}^\rho_1
    ,\overline{\xi}^\rho_2)}
    {\delta \rho_{i}(\overline{\xi}^{i})}
    \notag
    \\
    =&
    -
    \sum_{i_{1'}, i_{2'}}
    \int_{\overline{\xi}_{{2'}}^{i_{1'}},\overline{\xi}_{{2'}}^{i_{2'}}}
    \Gamma^{(2)-1}_{\lambda=0,\rho i_{1'}}[\vec{\rho}]
    (\overline{\xi}^{\rho}_1+\epsilon_\tau',\overline{\xi}^{i_{1'}}_{1'})
    \Gamma^{(2)-1}_{\lambda=0,\rho i_{2'}}[\vec{\rho}]
    (\overline{\xi}^{\rho}_2,\overline{\xi}^{i_{2'}}_{2'})
    \Gamma^{(3)}_{\lambda=0,i i_{1'} i_{2'}}[\vec{\rho}]
    (\overline{\xi}^{i},\overline{\xi}^{i_{1'}}_{1'},\overline{\xi}^{i_{2'}}_{2'})
    \notag
    \\
	&
	-
	\delta(\overline{\xi}^{i},\overline{\xi}_{12}^{\kappa})
	\kappa^*(\overline{\xi}_{12}^{\kappa^*})
	|_{\overline{\xi}_{12}^{\kappa^{(*)}}=({\overline{\xi}^\rho_1,\overline{\xi}^\rho_2})}
	-
	\delta(\overline{\xi}^{i},\overline{\xi}_{12}^{\kappa^*})
	\kappa(\overline{\xi}_{12}^{\kappa})
	|_{\overline{\xi}_{12}^{\kappa^{(*)}}=({\overline{\xi}^\rho_1,\overline{\xi}^\rho_2})}
	-
	\delta(\overline{\xi}^{i},\overline{\xi}^\rho_1)
	\delta_{a_1,a_2}
	\delta({\bf x}_1-{\bf x}_2).
\end{align}
Here, the following notation has been introduced:
\begin{align*}
	\Gamma^{(n)}_{\lambda, i_1\cdots i_n}[\vec{\rho}](\overline{\xi}^{i_1}_1,\cdots,\overline{\xi}^{i_n}_n)
	\coloneqq
	\frac{\delta^n \Gamma_\lambda[\vec{\rho}]}
	{\delta \rho_{i_1}(\overline{\xi}^{i_1}_{1})
	\cdots \delta \rho_{i_n}(\overline{\xi}^{i_n}_{n})},
\end{align*}
and $\Gamma^{(2)-1}_{\lambda, ij}[\vec{\rho}](\overline{\xi}^i,\overline{\xi'}^j)$ is equivalent to 
$(\delta^2\Gamma_\lambda/\delta\vec{\rho}\delta\vec{\rho})^{-1}_{ij}[\vec{\rho}](\overline{\xi}^i,\overline{\xi'}^j)$ 
defined in Eq.~\eqref{eq:gam2_inv}.
$\Gamma^{(n\geq 2)}_{\lambda=0}$ are 
expressed in terms of
 the density correlation functions of the non-interacting system,
which can be 
further reduced to simpler forms
using the Wick's theorem.
When the superfluidity is not considered~\cite{kem17a,yok18},
 $\Gamma^{(n\geq 2)}_{\lambda}$ are also written in terms of
the connected density correlation functions.
We are going to extend them so as to incorporate the anomalous density-correlation functions as
\begin{align}
	\Gamma^{(2)}_{\lambda,i_1 i_2}[\vec{\rho}]
	(\overline{\xi}^{i_1}_1, \overline{\xi}^{i_2}_2)
	=&
	G^{(2)-1}_{\lambda,i_1 i_2}[\vec{\rho}](\overline{\xi}^{i_1}_1, \overline{\xi}^{i_2}_2),
	\\
	\Gamma^{(3)}_{\lambda,i_1 i_2 i_3}[\vec{\rho}]
	(\overline{\xi}^{i_1}_1,\overline{\xi}^{i_2}_2,\overline{\xi}^{i_3}_3)
	=&
	-
	\sum_{i_4,i_5,i_6}
	\int_{\overline{\xi}^{i_4}_4,\overline{\xi}^{i_5}_5,\overline{\xi}^{i_6}_6}
	G^{(2)-1}_{\lambda,i_1 i_4}[\vec{\rho}](\overline{\xi}^{i_1}_1,\overline{\xi}^{i_4}_4)
	G^{(2)-1}_{\lambda,i_2 i_5}[\vec{\rho}](\overline{\xi}^{i_2}_2,\overline{\xi}^{i_5}_5)
	\notag
	\\
	&\times
	G^{(2)-1}_{\lambda,i_3 i_6}[\vec{\rho}](\overline{\xi}^{i_3}_3,\overline{\xi}^{i_6}_6)
	G^{(3)}_{\lambda,i_4 i_5 i_6}[\vec{\rho}](\overline{\xi}^{i_4}_4,\overline{\xi}^{i_5}_5,\overline{\xi}^{i_6}_6),
\end{align}
where $G^{(n)}_{\lambda, i_1\cdots i_n}(\overline{\xi}^{i_1}_{1},\cdots,\overline{\xi}^{i_n}_{n})$
is the connected correlation function given by
\begin{align*}
	G^{(n)}_{\lambda, i_1\cdots i_n}[\vec{\rho}](\overline{\xi}^{i_1}_1,\cdots,\overline{\xi}^{i_n}_n)
	=
	\frac{\delta^n W_\lambda[\vec{J}_{{\rm sup},\lambda}[\vec{\rho}]]}
	{\delta J_{i_1}(\overline{\xi}^{i_1}_{1})
	\cdots \delta J_{i_n}(\overline{\xi}^{i_n}_{n})},
\end{align*}
and $G^{(2)-1}_{\lambda=0,i_1 i_2}(\overline{\xi}^{i_1}_1,\overline{\xi}^{i_2}_2)$
satisfies
\begin{align*}
	\sum_{i_3} \int_{\overline{\xi}^{i_3}_3} 
	G^{(2)-1}_{\lambda, i_1 i_3}[\vec{\rho}](\overline{\xi}^{i_1}_{1},\overline{\xi}^{i_3}_{3})
	G^{(2)}_{\lambda, i_3 i_2}[\vec{\rho}](\overline{\xi}^{i_3}_{3},\overline{\xi}^{i_2}_{2})
	=
	\delta(\overline{\xi}^{i_1}_{1},\overline{\xi}^{i_2}_{2}).
\end{align*}
Using these relations, the last term in the right-hand side of Eq.~\eqref{eq:jcond_lo}
is rewritten as follows:
\begin{align}
	\label{eq:jcond_g}
	&
	\frac{1}{2}
	\int_{\overline{\xi}^{\rho}_1,	\overline{\xi}^{\rho}_2}
	\mathcal{U}_{\lambda=1}(\overline{\xi}^\rho_1,\overline{\xi}^\rho_2)
    \frac{\delta 
    G_{\rm x}^{(2)}[\vec{\rho}]
    (\overline{\xi}^\rho_1,\overline{\xi}^\rho_2)}
    {\delta \rho_{i}(\overline{\xi}^{i})}
    \notag
    \\
    =&
	\frac{1}{2}
	\int_{\overline{\xi}^{\rho}_1,	\overline{\xi}^{\rho}_2}
	\mathcal{U}_{\lambda=1}(\overline{\xi}^\rho_1,\overline{\xi}^\rho_2)
	\left[
    \sum_{i_{1'}}
    \int_{\overline{\xi}_{{1'}}^{i_{1'}}}
    G^{(2)-1}_{\lambda=0,i i_{1'}}[\vec{\rho}](\overline{\xi}^{i},\overline{\xi}^{i_{1'}}_{1'})
    G^{(3)}_{\lambda=0,i_{1'} \rho\rho}[\vec{\rho}]
    (\overline{\xi}^{i_{1'}}_{1'},\overline{\xi}^{\rho}_1+\epsilon_\tau',\overline{\xi}^{\rho}_2)
    \right.
    \notag
    \\
	&-
	\left.
	\delta(\overline{\xi}_{1}^{i},\overline{\xi}_{12}^{\kappa})
	\kappa^*(\overline{\xi}_{12}^{\kappa^*})
	|_{\overline{\xi}_{12}^{\kappa^{(*)}}=({\overline{\xi}^\rho_1,\overline{\xi}^\rho_2})}
	-
	\delta(\overline{\xi}^{i},\overline{\xi}_{12}^{\kappa^*})
	\kappa(\overline{\xi}_{12}^{\kappa})
	|_{\overline{\xi}_{12}^{\kappa^{(*)}}=({\overline{\xi}^\rho_1,\overline{\xi}^\rho_2})}
	-
	\delta(\overline{\xi}^{i},\overline{\xi}^\rho_1)
	\delta_{a_1,a_2}
	\delta({\bf x}_1-{\bf x}_2)
    \right]
    \notag
    \\
    =&
	\frac{1}{2}
    \sum_{i_{1'}}
    \int_{\overline{\xi}_{{1'}}^{i_{1'}}}
    G^{(2)-1}_{\lambda=0,i i_{1'}}[\vec{\rho}](\overline{\xi}^{i},\overline{\xi}^{i_{1'}}_{1'})
	\int_{\overline{\xi}^{\rho}_1,	\overline{\xi}^{\rho}_2}
	\mathcal{U}_{\lambda=1}(\overline{\xi}^\rho_1,\overline{\xi}^\rho_2)
    \notag
    \\
	&\times
	\left[
    G^{(3)}_{\lambda=0,i_{1'} \rho\rho}[\vec{\rho}]
    (\overline{\xi}^{i_{1'}}_{1'},\overline{\xi}^{\rho}_1+\epsilon_\tau',\overline{\xi}^{\rho}_2)
    -
	G^{(2)}_{\lambda=0,i_{1'} \kappa}[\vec{\rho}]
	(\overline{\xi}^{i_{1'}}_{1'},\overline{\xi}^{\kappa}_{12})
	\kappa^*(\overline{\xi}_{12}^{\kappa^*})
	|_{\overline{\xi}_{12}^{\kappa^{(*)}}=({\overline{\xi}^\rho_1,\overline{\xi}^\rho_2})}
	\right.
	\notag
	\\
	&
	\left.
	-
	G^{(2)}_{\lambda=0,i_{1'} \kappa^*}[\vec{\rho}]
	(\overline{\xi}^{i_{1'}}_{1'},\overline{\xi}^{\kappa^*}_{12})
	\kappa(\overline{\xi}_{12}^{\kappa})
	|_{\overline{\xi}_{12}^{\kappa^{(*)}}=({\overline{\xi}^\rho_1,\overline{\xi}^\rho_2})}
	-
	G^{(2)}_{\lambda=0,i_{1'} \rho}
	(\overline{\xi}^{i_{1'}}_{1'},\overline{\xi}^\rho_2)
	\delta_{a_1,a_2}
	\delta({\bf x}_1-{\bf x}_2)
    \right].
\end{align}
By use of the Wick's theorem, 
$G^{(3)}_{\lambda=0,i_{1'} \rho\rho}[\vec{\rho}]
    (\overline{\xi}^{i_{1'}}_{1'},\overline{\xi}^{\rho}_1+\epsilon_\tau',\overline{\xi}^{\rho}_2)$
is evaluated as follows:
\begin{align*}
    &G^{(3)}_{\lambda=0,i_{1'}\rho\rho}[\vec{\rho}]
    (\overline{\xi}_{{1'}}^{i_{1'}},
    \overline{\xi}_1^\rho+\epsilon'_\tau,\overline{\xi}_2^\rho)
    \\
    =
    &
    \langle
    \psi^*(\xi_1)\psi^*(\xi_{2}+\epsilon'_\tau)
    \rangle_{\vec{\rho}}
    \left(
    \langle
    \psi(\xi_2)\psi(\xi_1+\epsilon'_\tau)
    \hat{\rho}_{i_{1'}}(\overline{\xi}^{i_{1'}}_{1'})
    \rangle_{\vec{\rho}}
    -
    \langle
    \psi(\xi_{2})
    \psi(\xi_1+\epsilon'_\tau)
    \rangle_{\vec{\rho}}
    \langle
    \hat{\rho}_{i_{1'}}(\overline{\xi}_{{1'}}^{i_{1'}})
    \rangle_{\vec{\rho}}
    \right)
    \\
    &
    +
    \langle
    \psi(\xi_{2})\psi(\xi_1+\epsilon'_\tau)
    \rangle_{\vec{\rho}}
    \left(
    \langle
    \psi^*(\xi_1+\epsilon'_\tau)\psi^*(\xi_{2})
    \hat{\rho}_{i_{1'}}(\overline{\xi}^{i_{1'}}_{1'})
    \rangle_{\vec{\rho}}
    -
    \langle
    \psi^*(\xi_1+\epsilon'_\tau)\psi^*(\xi_{2})
    \rangle_{\vec{\rho}}
    \langle
    \hat{\rho}_{i_{1'}}(\overline{\xi}_{{1'}}^{i_{1'}})
    \rangle_{\vec{\rho}}
    \right)
    \\
    &
    -
    \langle
    \psi^*(\xi_{2})\psi(\xi_1+\epsilon'_\tau)
    \rangle_{\vec{\rho}}
    \left(
    \langle
    \psi^*(\xi_1+\epsilon'_\tau)\psi(\xi_{2})
    \hat{\rho}_{i_{1'}}(\overline{\xi}^{i_{1'}}_{1'})
    \rangle_{\vec{\rho}}
    -
    \langle
    \psi^*(\xi_1+\epsilon'_\tau)\psi(\xi_{2})
    \rangle_{\vec{\rho}}
    \langle
    \hat{\rho}_{i_{1'}}(\overline{\xi}_{{1'}}^{i_{1'}})
    \rangle_{\vec{\rho}}
    \right)
    \\
    &
    -
    \langle
    \psi^*(\xi_1+\epsilon'_\tau)\psi(\xi_{2})
    \rangle_{\vec{\rho}}
    \left(
    \langle
    \psi^*(\xi_{2})\psi(\xi_1+\epsilon'_\tau)
    \hat{\rho}_{i_{1'}}(\overline{\xi}_{{1'}}^{i_{1'}})
    \rangle_{\vec{\rho}}
    -
    \langle
    \psi^*(\xi_{2})\psi(\xi_1+\epsilon'_\tau)
    \rangle_{\vec{\rho}}
    \langle
    \hat{\rho}_{i_{1'}}(\overline{\xi}_{{1'}}^{i_{1'}})
    \rangle_{\vec{\rho}}
    \right)
    \\
    =
    &
    \kappa^*(\overline{\xi}^{\kappa^*}_{12})
    G^{(2)}_{\lambda=0,i_{1'} \kappa}[\vec{\rho}](\overline{\xi}^{i_{1'}}_{1'},\overline{\xi}_{12}^\kappa)
    |_{\overline{\xi}^{\kappa^{(*)}}=(\overline{\xi}^\rho_1,\overline{\xi}^\rho_2)}
    +
    \kappa(\overline{\xi}_{12}^{\kappa})
    G^{(2)}_{\lambda=0,i_{1'} \kappa^*}[\vec{\rho}](\overline{\xi}^{i_{1'}}_{1'},\overline{\xi}^{\kappa^*}_{12})
    |_{\overline{\xi}^{\kappa^{(*)}}=(\overline{\xi}^\rho_1,\overline{\xi}^\rho_2)}
    \\
    &
    -
    \langle
    \psi^*(\xi_{2})\psi(\xi_1+\epsilon'_\tau)
    \rangle_{\vec{\rho}}
    \left(
    \langle
    \psi^*(\xi_1+\epsilon'_\tau)\psi(\xi_{2})
    \hat{\rho}_{i_{1'}}(\overline{\xi}^{i_{1'}}_{1'})
    \rangle_{\vec{\rho}}
    -
    \langle
    \psi^*(\xi_1+\epsilon'_\tau)\psi(\xi_{2})
    \rangle_{\vec{\rho}}
    \langle
    \hat{\rho}_{i_{1'}}(\overline{\xi}_{{1'}}^{i_{1'}})
    \rangle_{\vec{\rho}}
    \right)
    \\
    &
    -
    \langle
    \psi^*(\xi_1+\epsilon'_\tau)\psi(\xi_{2})
    \rangle_{\vec{\rho}}
    \left(
    \langle
    \psi^*(\xi_{2})\psi(\xi_1+\epsilon'_\tau)
    \hat{\rho}_{i_{1'}}(\overline{\xi}^{i_{1'}}_{1'})
    \rangle_{\vec{\rho}}
    -
    \langle
    \psi^*(\xi_{2})\psi(\xi_1+\epsilon'_\tau)
    \rangle_{\vec{\rho}}
    \langle
    \hat{\rho}_{i_{1'}}(\overline{\xi}_{{1'}}^{i_{1'}})
    \rangle_{\vec{\rho}}
    \right).
\end{align*}
Here, we have introduced
\begin{align}
	\label{eq:ave_o_r}
	\langle\mathcal{O}[\psi,\psi^*]\rangle_{\vec{\rho}}:=
\langle\mathcal{O}[\psi,\psi^*]\rangle_{\lambda=0,\vec{J}_{\rm sup,\lambda=0}[\vec{\rho}]},
\end{align}
in which the right-hand side is defined by Eq.~\eqref{eq:ave_def}
for a functional $\mathcal{O}[\psi,\psi^*]$.
Then the integral with respect to $\overline{\xi}^{\rho}_1$ and 
$\overline{\xi}^{\rho}_2$
in Eq.~\eqref{eq:jcond_g} is evaluated as follows:
\begin{align}
	\label{eq:ug3int}
	&\int_{\overline{\xi}^{\rho}_1,	\overline{\xi}^{\rho}_2}
	\mathcal{U}_{\lambda=1}(\overline{\xi}^\rho_1,\overline{\xi}^\rho_2)
	\left[
    G^{(3)}_{\lambda=0,i_{1'} \rho\rho}[\vec{\rho}]
    (\overline{\xi}^{i_{1'}}_{1'},\overline{\xi}^{\rho}_1+\epsilon_\tau',\overline{\xi}^{\rho}_2)
    \right.
    \notag
    \\
    &-
    \left.
	G^{(2)}_{\lambda=0,i_{1'} \kappa}[\vec{\rho}]
	(\overline{\xi}^{i_{1'}}_{1'},\overline{\xi}^{\kappa}_{12})
	\kappa^*(\overline{\xi}_{12}^{\kappa^*})
	|_{\overline{\xi}_{12}^{\kappa^{(*)}}=({\overline{\xi}^\rho_1,\overline{\xi}^\rho_2})}
	-
	G^{(2)}_{\lambda=0,i_{1'} \kappa^*}[\vec{\rho}]
	(\overline{\xi}^{i_{1'}}_{1'},\overline{\xi}^{\kappa^*}_{12})
	\kappa(\overline{\xi}_{12}^{\kappa})
	|_{\overline{\xi}_{12}^{\kappa^{(*)}}=({\overline{\xi}^\rho_1,\overline{\xi}^\rho_2})}
	\right.
	\notag
	\\
	&-
	\left.
	G^{(2)}_{\lambda=0,i_{1'} \rho}[\vec{\rho}]
	(\overline{\xi}^{i_{1'}}_{1'},\overline{\xi}^\rho_1)
	\delta_{a_1,a_2}
	\delta({\bf x}_1-{\bf x}_2)
    \right]
    \notag
    \\
    =&
    -
    2\int_{\overline{\xi}_{1}^\rho,\overline{\xi}_2^\rho}
	\mathcal{U}_{\lambda=1}(\overline{\xi}_1^\rho,\overline{\xi}_2^\rho)
    \langle
    \psi^*(\xi_{1}+\epsilon'_\tau)\psi(\xi_2)
    \rangle_{\vec{\rho}}
    \notag
    \\
    &\times\left(
    \langle
    \psi^*(\xi_2)\psi(\xi_{1}+\epsilon'_\tau)
    \hat{\rho}_{i_{1'}}(\overline{\xi}^{i_{1'}}_{1'})
    \rangle_{\vec{\rho}}
    -
    \langle
    \psi^*(\xi_2)\psi(\xi_{1}+\epsilon'_\tau)
    \rangle_{\vec{\rho}}
    \langle
    \hat{\rho}_{i_{1'}}(\overline{\xi}^{i_{1'}}_{1'})
    \rangle_{\vec{\rho}}
    \right),
\end{align}
where we have used the equal-time commutation relation:
\begin{align*}
	\mathcal{U}_{\lambda=1}(\overline{\xi}_1^\rho,\overline{\xi}_2^\rho)
    \langle
    \psi^*(\xi_1+\epsilon'_\tau)\psi(\xi_{2})
    \rangle_{\vec{\rho}}
    =&
    \delta(\tau_1-\tau_2)U_{\lambda,a_1 a_2}({\bf x}_1-{\bf x}_2)
    \langle
    \psi^*(\xi_1+\epsilon'_\tau)\psi(\xi_{2})
    \rangle_{\vec{\rho}}
    \\
    =&
    \delta(\tau_1-\tau_2)U_{\lambda,a_1 a_2}({\bf x}_1-{\bf x}_2)
    \left(
    \langle
    \psi^*(\xi_{1})\psi(\xi_{2}+\epsilon'_\tau)
    \rangle_{\vec{\rho}}
    -
    \delta_{a_1,a_2}\delta({\bf x}_1-{\bf x}_2)
    \right).
\end{align*}
Substituting Eq.~\eqref{eq:ug3int} into Eq.~\eqref{eq:jcond_g},
we have
\begin{align}
	\label{eq:jcond_sol}
	\mathcal{V}_{\rm KS}^i[\vec{\rho}](\overline{\xi}^i)
    =
    &
    \delta_{i,\rho}
    \left(
    \mathcal{V}(\overline{\xi}^\rho)
    +
	\int_{\overline{\xi}^\rho_1}
	\mathcal{U}_{\lambda=1}(\overline{\xi}^\rho,\overline{\xi}^\rho_1)
	\rho(\overline{\xi}^\rho_1)
	\right)
    +
    \frac{\delta_{i,\kappa}}{2}
	\overline{\mathcal{U}}_{\lambda=1}(\overline{\xi}^{\kappa})
	\kappa^*(\overline{\xi}^{\kappa})
    +
    \frac{\delta_{i,\kappa^*}}{2}
	\overline{\mathcal{U}}_{\lambda=1}(\overline{\xi}^{\kappa^*})
	\kappa(\overline{\xi}^{\kappa^*})
	\notag
	\\
	&
	-\sum_{i_{1'}}
    \int_{\overline{\xi}_{{1'}}^{i_{1'}}}
    G^{(2)-1}_{\lambda=0,i i_{1'}}[\vec{\rho}](\overline{\xi}^{i},\overline{\xi}^{i_{1'}}_{1'})
    \int_{\overline{\xi}_{1}^\rho,\overline{\xi}_2^\rho}
	\mathcal{U}_{\lambda=1}(\overline{\xi}_1^\rho,\overline{\xi}_2^\rho)
    \langle
    \psi^*(\xi_{1}+\epsilon'_\tau)\psi(\xi_2)
    \rangle_{\vec{\rho}}
    \notag
    \\
    &\times\left(
    \langle
    \psi^*(\xi_2)\psi(\xi_{1}+\epsilon'_\tau)
    \hat{\rho}_{i_{1'}}(\overline{\xi}^{i_{1'}}_{1'})
    \rangle_{\vec{\rho}}
    -
    \langle
    \psi^*(\xi_2)\psi(\xi_{1}+\epsilon'_\tau)
    \rangle_{\vec{\rho}}
    \langle
    \hat{\rho}_{i_{1'}}(\overline{\xi}^{i_{1'}}_{1'})
    \rangle_{\vec{\rho}}
    \right).
\end{align}
Although the last term in the right-hand side may not be possible
to represent in terms of the density correlation functions in general,
it can be further simplified for an arbitrary $\mathcal{U}_{\lambda}$.
In the next subsection, we 
shall discuss the case with short-range interactions 
and then show that the last term of Eq.~\eqref{eq:jcond_sol} is 
approximately represented 
in terms of the density correlation functions. 

Finally, let us discuss the Helmholtz free energy.
Applying the approximation 
$G^{(2)}_{\rm xc,\lambda}\approx G^{(2)}_{\rm x}$
to Eq.~\eqref{eq:fave_exp},
we have 
\begin{align}
	\label{eq:fave_lo}
	F_{\rm H}[\vec{\rho}_{\rm ave}]
	=
	&
	\frac{\left.\Gamma_{\lambda=0}\right|_{\mathcal{V}=0}[\vec{\rho}_{\rm ave}]}{\beta}
	+
	\frac{1}{\beta}\int_{\xi} \mathcal{V}(\xi)\rho_{\rm ave}(\xi)
	\notag
	\\
	&
	+
	\frac{1}{2\beta}
	\int_{\overline{\xi}^\rho_1,\overline{\xi}^\rho_2}
	\mathcal{U}_{\lambda=1}(\overline{\xi}^\rho_1,\overline{\xi}^\rho_2)
	\left[
	\rho_{\rm ave}(\overline{\xi}^\rho_1)\rho_{\rm ave}(\overline{\xi}^\rho_2)
	+
	\kappa^*_{\rm ave}(\overline{\xi}_{12}^{\kappa^*})
	\kappa_{\rm ave}(\overline{\xi}_{12}^{\kappa})
	|_{\overline{\xi}_{12}^{\kappa^{(*)}}=({\overline{\xi}^\rho_1,\overline{\xi}^\rho_2})}
	\right]
	\notag
	\\
	&
	+
	\frac{1}{2\beta}
	\int_{\overline{\xi}^\rho_1,\overline{\xi}^\rho_2}
	\mathcal{U}_{\lambda=1}(\overline{\xi}^\rho_1,\overline{\xi}^\rho_2)
	G^{(2)}_{{\rm x}}[\vec{\rho}_{\rm ave}]({\overline{\xi}^\rho_1,\overline{\xi}^\rho_2}).
\end{align}
From Eqs.~\eqref{eq:gamma_decomp} and \eqref{eq:g2x}, 
$G^{(2)}_{{\rm x}}[\vec{\rho}]({\overline{\xi}^\rho_1,\overline{\xi}^\rho_2})$ is rewritten as
\begin{align*}
	G^{(2)}_{{\rm x}}[\vec{\rho}]({\overline{\xi}^\rho_1,\overline{\xi}^\rho_2})
	=
	-
	\langle
	\psi^*(\xi_1+\epsilon_\tau+\epsilon'_\tau)
	\psi(\xi_2)
	\rangle_{\vec{\rho}}
	\langle
	\psi^*(\xi_2+\epsilon_\tau)
	\psi(\xi_1+\epsilon'_\tau)
	\rangle_{\vec{\rho}}
	-
	\rho(\overline{\xi}^\rho_1)
	\delta_{a_1,a_2}
	\delta({\bf x}_1-{\bf x}_2).
\end{align*}
Then the integral with respect to $\overline{\xi}^{\rho}_1$ and $\overline{\xi}^{\rho}_2$
in the third line of Eq.~\eqref{eq:fave_lo} is evaluated as follows:
\begin{align}
	\label{eq:ug2int}
	\int_{\overline{\xi}^\rho_1,\overline{\xi}^\rho_2}
	\mathcal{U}_{\lambda=1}(\overline{\xi}^\rho_1,\overline{\xi}^\rho_2)
	&
	G^{(2)}_{{\rm x}}[\vec{\rho}_{\rm ave}]({\overline{\xi}^\rho_1,\overline{\xi}^\rho_2})
	\notag \\
	&
	=
	-
	\int_{\overline{\xi}^\rho_1,\overline{\xi}^\rho_2}
	\mathcal{U}_{\lambda=1}(\overline{\xi}^\rho_1,\overline{\xi}^\rho_2)
	\langle
	\psi^*(\xi_1+\epsilon_\tau+\epsilon'_\tau)
	\psi(\xi_2)
	\rangle_{\vec{\rho}_{\rm ave}}
	\langle
	\psi^*(\xi_2+\epsilon_\tau+\epsilon'_\tau)
	\psi(\xi_1)
	\rangle_{\vec{\rho}_{\rm ave}},
\end{align}
where we have used the equal-time commutation relation:
\begin{align*}
	\mathcal{U}_{\lambda=1}(\overline{\xi}^\rho_1,\overline{\xi}^\rho_2)
	&
	\langle
	\psi^*(\xi_2+\epsilon_\tau)
	\psi(\xi_1+\epsilon'_\tau)
	\rangle_{\vec{\rho}}
	\\
	&=
	\delta(\tau_1-\tau_2)U_{\lambda,a_1 a_2}({\bf x}_1-{\bf x}_2)
    \langle
    \psi^*(\xi_2+\epsilon_\tau)\psi(\xi_{1}+\epsilon'_\tau)
    \rangle_{\vec{\rho}}
    \\
    &=
    \delta(\tau_1-\tau_2)U_{\lambda,a_1 a_2}({\bf x}_1-{\bf x}_2)
    \left(
    \langle
    \psi^*(\xi_{2}+\epsilon_\tau+\epsilon'_\tau)\psi(\xi_{1})
    \rangle_{\vec{\rho}}
    -
    \delta_{a_1,a_2}\delta({\bf x}_1-{\bf x}_2)
    \right).
\end{align*}
Substituting Eq.~\eqref{eq:ug2int} into Eq.~\eqref{eq:fave_lo},
we have
\begin{align}
	\label{eq:fave_lo2}
	F_{\rm H}&[\vec{\rho}_{\rm ave}]
	=
	\frac{\left.\Gamma_{\lambda=0}\right|_{\mathcal{V}=0}[\vec{\rho}_{\rm ave}]}{\beta}
	+
	\frac{1}{\beta}\int_{\xi} \mathcal{V}(\xi)\rho_{\rm ave}(\xi)
	\notag
	\\
	&
	+
	\frac{1}{2\beta}
	\int_{\overline{\xi}^\rho_1,\overline{\xi}^\rho_2}
	\mathcal{U}_{\lambda=1}(\overline{\xi}^\rho_1,\overline{\xi}^\rho_2)
	\left[
	\rho_{\rm ave}(\overline{\xi}^\rho_1)\rho_{\rm ave}(\overline{\xi}^\rho_2)
	-
	\langle
	\psi^*(\xi_1+\epsilon_\tau+\epsilon'_\tau)
	\psi(\xi_2)
	\rangle_{\vec{\rho}_{\rm ave}}
	\langle
	\psi^*(\xi_2+\epsilon_\tau+\epsilon'_\tau)
	\psi(\xi_1)
	\rangle_{\vec{\rho}_{\rm ave}}
	\right]
	\notag
	\\
	&
	+
	\frac{1}{2\beta}
	\int_{\overline{\xi}^\rho_1,\overline{\xi}^\rho_2}
	\mathcal{U}_{\lambda=1}(\overline{\xi}^\rho_1,\overline{\xi}^\rho_2)
	\kappa^*_{\rm ave}(\overline{\xi}_{12}^{\kappa^*})
	\kappa_{\rm ave}(\overline{\xi}_{12}^{\kappa})
	|_{\overline{\xi}_{12}^{\kappa^{(*)}}=({\overline{\xi}^\rho_1,\overline{\xi}^\rho_2})}.
\end{align}
At the zero temperature limit $\beta \to \infty$, it gives the ground-state energy:
\begin{align*}
	&E_{\rm gs}
	=
	T[\vec{\rho}_{\rm gs}]
	+
	\lim_{\beta\to\infty}
	\frac{1}{\beta}\int_{\xi} \mathcal{V}(\xi)\rho_{\rm ave}(\xi)
	\notag
	\\
	&
	+
	\lim_{\beta\to\infty}
	\frac{1}{2\beta}
	\int_{\overline{\xi}^\rho_1,\overline{\xi}^\rho_2}
	\mathcal{U}_{\lambda=1}(\overline{\xi}^\rho_1,\overline{\xi}^\rho_2)
	\left[
	\rho_{\rm ave}(\overline{\xi}^\rho_1)\rho_{\rm ave}(\overline{\xi}^\rho_2)
	-
	\langle
	\psi^*(\xi_1+\epsilon_\tau+\epsilon'_\tau)
	\psi(\xi_2)
	\rangle_{\vec{\rho}_{\rm ave}}
	\langle
	\psi^*(\xi_2+\epsilon_\tau+\epsilon'_\tau)
	\psi(\xi_1)
	\rangle_{\vec{\rho}_{\rm ave}}
	\right]
	\notag
	\\
	&
	+
	\lim_{\beta\to\infty}
	\frac{1}{2\beta}
	\int_{\overline{\xi}^\rho_1,\overline{\xi}^\rho_2}
	\mathcal{U}_{\lambda=1}(\overline{\xi}^\rho_1,\overline{\xi}^\rho_2)
	\kappa^*_{\rm ave}(\overline{\xi}_{12}^{\kappa^*})
	\kappa_{\rm ave}(\overline{\xi}_{12}^{\kappa})
	|_{\overline{\xi}_{12}^{\kappa^{(*)}}=({\overline{\xi}^\rho_1,\overline{\xi}^\rho_2})}.
\end{align*}
The third term in the right-hand side
is interpreted as the Hartree--Fock energy; 
The fourth term is the contribution from the pairing condensate.
This formula of the ground-state energy
is identical to that given in the Hartree--Fock--Bogoliubov approximation~\cite{rin80,deg99,tin04}.

\subsection{The case of short-range interactions in the weak coupling}

Let us take
 the case where the inter-particle interaction is short-range 
in the weak coupling limit of the paring force as in the BCS theory \cite{BCS57}.
Then it will be found that Eq.~\eqref{eq:jcond_sol} 
is reduced to a simple and familiar form.

In the case of short-range interactions,
the integral in the last term of Eq.~\eqref{eq:jcond_sol} is dominated by
the contribution 
from the region where ${\bf x}_2$ is close to ${\bf x}_1$. 
Therefore, we can make the following approximation for short-range 
interactions that holds exactly for the zero-range interaction:
\begin{align}
&
	\label{eq:exapprox1}
    \int_{\overline{\xi}_{1}^\rho,\overline{\xi}_2^\rho}
	\mathcal{U}_{\lambda=1}(\overline{\xi}_1^\rho,\overline{\xi}_2^\rho)
    \langle
    \psi^*(\xi_{1}+\epsilon'_\tau)\psi(\xi_2)
    \rangle_{\vec{\rho}}
    \left(
    \langle
    \psi^*(\xi_2)\psi(\xi_{1}+\epsilon'_\tau)
    \hat{\rho}_{i_{1'}}(\overline{\xi}^{i_{1'}}_{1'})
    \rangle_{\vec{\rho}}
    -
    \langle
    \psi^*(\xi_2)\psi(\xi_{1}+\epsilon'_\tau)
    \rangle_{\vec{\rho}}
    \langle
    \hat{\rho}_{i_{1'}}(\overline{\xi}^{i_{1'}}_{1'})
    \rangle_{\vec{\rho}}
    \right)
    \notag
    \\
    \approx&
    \int_{\overline{\xi}_{1}^\rho=(\tau_1,{\bf x}_1,a_1),\overline{\xi}_2^\rho=(\tau_2,{\bf x}_2,a_2)}
	\mathcal{U}_{\lambda=1}(\overline{\xi}_1^\rho,\overline{\xi}_2^\rho)
	\rho_{a_1 a_2}(\tau_2,{\bf x}_2)
    \notag
    \\
    &\times
    \left(
    \langle
    \psi^*(\tau_1,{\bf x}_1,a_2)
    \psi(\tau_1+\epsilon'_\tau,{\bf x}_1,a_1)
    \hat{\rho}_{i_{1'}}(\overline{\xi}^{i_{1'}}_{1'})
    \rangle_{\vec{\rho}}
    -
    \langle
    \psi^*(\tau_1,{\bf x}_1,a_2)
    \psi(\tau_1+\epsilon'_\tau,{\bf x}_1,a_1)
    \rangle_{\vec{\rho}}
    \langle
    \hat{\rho}_{i_{1'}}(\overline{\xi}^{i_{1'}}_{1'})
    \rangle_{\vec{\rho}}
    \right),
\end{align}
where we have used
\[
\mathcal{U}_{\lambda=1}(\overline{\xi}_1^\rho:=(\tau_1,{\bf x}_1,a_1),\overline{\xi}_2^\rho:=(\tau_2,{\bf x}_2,a_2))
= \delta(\tau_1-\tau_2)\mathcal{U}_{\lambda=1}(\tau_1,{\bf x}_1,a_1;\, \tau_1,{\bf x}_2,a_2),
\]
and introduced $\rho_{a_1 a_2}(\tau,{\bf x})=
    \langle
    \psi^*(\tau+\epsilon'_\tau,{\bf x},a_1)
    \psi(\tau,{\bf x},a_2)
    \rangle_{\vec{\rho}}$.

We evaluate $\rho_{a_1 a_2}(\tau,{\bf x})$ in the weak coupling limit leading to
 small $\kappa$ in comparison with the particle density $\rho$.
In the non-interacting case $\lambda=0$,
the vanishing external field $J_{\rm sup,\lambda=0}^{\kappa^{(*)}}[\vec{\rho}]=0$
gives $\kappa^{(*)}=0$.
Conversely, we have $\lim_{\kappa,\kappa^{*}\to 0}J_{\rm sup,\lambda=0}^{\kappa^{(*)}}[\vec{\rho}]=0$
since $\kappa^{(*)}$ uniquely determines the external field.
Therefore, by use of Eqs.~\eqref{eq:ave_o_r} and \eqref{eq:ave_def}, $\rho_{a_1 a_2}(\tau,{\bf x})$
is approximated in the weak coupling as
\begin{align}
\rho_{a_1 a_2}(\tau,{\bf x})\,&\approx&
	\notag
	\\
	&&
\hspace{-1cm}	\frac{1}{Z_{\lambda=0}[(J_{\rm sup,\lambda=0}^\rho[\vec{\rho}],0,0)]}
    \int \mathcal{D}\psi \mathcal{D}\psi^*
    \psi^*(\tau+\epsilon'_\tau,{\bf x},a_1)
    \psi(\tau,{\bf x},a_2)
    e^{-S_{\lambda=0}[\psi, \psi^*]
    +
    \int_{\xi} J_{\rm sup,\lambda=0}^\rho[\vec{\rho}](\xi)
    \hat{\rho}(\xi)}.
\end{align}
Notice that there is no term causing a mixing 
between $\psi^{(*)}(\tau,{\bf x},a)$ 
with different $a$'s
in the exponent $S_{\lambda=0}[\psi, \psi^*]
    -\int_{\xi} J_{\rm sup,\lambda=0}^\rho[\vec{\rho}](\xi)
    \hat{\rho}(\xi)$ any more.
Thus we have $\rho_{a_1 a_2}(\tau,{\bf x})\simeq 0$
for $a_1\neq a_2$ in the weak coupling limit.
In other words, $\rho_{a_1 a_2}(\tau,{\bf x})$ with $a_1\neq a_2$
is negligible in comparison with the diagonal component
$\rho_{a_1 a_1}(\tau,{\bf x})=\rho(\tau,{\bf x},a_1)$
in the weak coupling limit or small $\kappa^{(*)}$.
Thus, we can make the approximation for $\rho_{a_1 a_2}(\tau,{\bf x})$ as follows:
\begin{align}
	\label{eq:approx_smallk}
	\rho_{a_1 a_2}(\tau,{\bf x})
	\approx
	\delta_{a_1,a_2} \rho(\tau,{\bf x},a_1).
\end{align}
Then, Eq.~\eqref{eq:exapprox1} is 
reduced as follows,
\begin{align}
&
	\label{eq:exapprox}
    \int_{\overline{\xi}_{1}^\rho,\overline{\xi}_2^\rho}
	\mathcal{U}_{\lambda=1}(\overline{\xi}_1^\rho,\overline{\xi}_2^\rho)
    \langle
    \psi^*(\xi_{1}+\epsilon'_\tau)\psi(\xi_2)
    \rangle_{\vec{\rho}}
    \left(
    \langle
    \psi^*(\xi_2)\psi(\xi_{1}+\epsilon'_\tau)
    \hat{\rho}_{i_{1'}}(\overline{\xi}^{i_{1'}}_{1'})
    \rangle_{\vec{\rho}}
    -
    \langle
    \psi^*(\xi_2)\psi(\xi_{1}+\epsilon'_\tau)
    \rangle_{\vec{\rho}}
    \langle
    \hat{\rho}_{i_{1'}}(\overline{\xi}^{i_{1'}}_{1'})
    \rangle_{\vec{\rho}}
    \right)
    \notag
    \\
    \approx&
    \int_{\overline{\xi}_{1}^\rho=(\tau_1,{\bf x}_1,a_1),\overline{\xi}_2^\rho=(\tau_2,{\bf x}_2,a_2)}
	\mathcal{U}_{\lambda=1}(\overline{\xi}_1^\rho,\overline{\xi}_2^\rho)
	\delta_{a_1 a_2}\rho(\tau_2,{\bf x}_2,a_2)
    \notag
    \\
    &\times
    \left(
    \langle
    \psi^*(\tau_1,{\bf x}_1,a_1)
    \psi(\tau_1+\epsilon'_\tau,{\bf x}_1,a_1)
    \hat{\rho}_{i_{1'}}(\overline{\xi}^{i_{1'}}_{1'})
    \rangle_{\vec{\rho}}
    -
    \langle
    \psi^*(\tau_1,{\bf x}_1,a_1)
    \psi(\tau_1+\epsilon'_\tau,{\bf x}_1,a_1)
    \rangle_{\vec{\rho}}
    \langle
    \hat{\rho}_{i_{1'}}(\overline{\xi}^{i_{1'}}_{1'})
    \rangle_{\vec{\rho}}
    \right)
    \notag
    \\
    =&
    \int_{\overline{\xi}_1^\rho,\overline{\xi}_2^\rho}
	\mathcal{U}_{\lambda=1}(\overline{\xi}_1^\rho,\overline{\xi}_2^\rho)
    \delta_{a_1,a_2}
    \rho(\overline{\xi}_2^\rho)
    G^{(2)}_{\lambda=0,\rho i_{1'}}[\vec{\rho}](\overline{\xi}_1^\rho, \overline{\xi}^{i_{1'}}_{1'}).
\end{align}
Applying this approximation to Eq.~\eqref{eq:jcond_sol},
we arrive at the following expression 
of the basic equation for the KS theory 
for the case of short-range interactions 
in the weak coupling as
\begin{align}
	\label{eq:jcond_final}
	\mathcal{V}_{\rm KS}^i[\vec{\rho}](\overline{\xi}^i)
    =
    &
    \delta_{i,\rho}
    \left(
    \mathcal{V}(\overline{\xi}^\rho)
    +
	\int_{\overline{\xi}^\rho_1}
	(1-\delta_{a,a_1})
	\mathcal{U}_{\lambda=1}(\overline{\xi}^\rho,\overline{\xi}^\rho_1)
	\rho(\overline{\xi}^\rho_1)
	\right)
	\notag
	\\
    &+
    \frac{\delta_{i,\kappa}}{2}
	\overline{\mathcal{U}}_{\lambda=1}(\overline{\xi}^{\kappa})
	\kappa^*(\overline{\xi}^{\kappa})
    +
    \frac{\delta_{i,\kappa^*}}{2}
	\overline{\mathcal{U}}_{\lambda=1}(\overline{\xi}^{\kappa^*})
	\kappa(\overline{\xi}^{\kappa^*}).
\end{align}

\subsection{Homogeneous systems} \label{sec:homo}

In this subsection, we are going to discuss a homogeneous system
with $\mathcal{V}(\xi)=0$ in Eq.~\eqref{eq:action},
and show that Eq.~\eqref{eq:jcond_final} together with Eq.~\eqref{eq:ks_vari} is reduced to the well-known gap equation in the BCS theory \cite{mah00}.
For simplicity, we consider fermions with spin but not any other internal degrees of freedom:
we assume that the spin is saturated in the system, and accordingly
the chemical potential $\mu$ has no spin dependence.
We also assume that the two-body interaction is 
a spin-independent short-range one $U_{ab}({\bf x})=U({\bf x})$ 
and consider the weak-coupling limit
so that the approximated equation Eq.~\eqref{eq:exapprox} is applicable.
Although we treat
 the spin-singlet condensate in the present investigation exclusively in the 
present article,
the extension to
the spin-triplet case is straightforward by considering the vector pairing densities.

Here, we introduce the following notation: 
$\mathcal{V}_{\rm KS}^i(\overline{\xi}^i):=\mathcal{V}_{\rm KS}^i[\vec{\rho}_{\rm ave}](\overline{\xi}^i)$.
Owing to the translational symmetry, the coordinate dependence of 
$\mathcal{V}_{\rm KS}^\rho	(\overline{\xi}^\rho)$ is simplified as follows:
\begin{align*}
	\mathcal{V}_{\rm KS}^\rho
	(\overline{\xi}^\rho:=(\tau,{\bf x},a))
    =&
	\mathcal{V}_{\rm KS}^\rho(0,{\bf 0},a)
    \eqqcolon
	\mathcal{V}_{{\rm KS},a}^\rho,
    \\
    \mathcal{V}_{\rm KS}^{\kappa^{(*)}}
    (\xi:=(\tau,{\bf x},a),\xi':=(\tau',{\bf x}',a'))
    =&
    \mathcal{V}_{\rm KS}^{\kappa^{(*)}}
    (\tau-\tau',{\bf x}-{\bf x}',a,0,{\bf 0},a')
    \eqqcolon
    \mathcal{V}_{{\rm KS},aa'}^{\kappa^{(*)}}
    (\tau-\tau',{\bf x}-{\bf x}').
\end{align*}
Similarly, we have
\begin{align*}
    \rho_{\rm ave}(\xi:=(\tau,{\bf x},a))
    =&
    \rho_{\rm ave}
    (\tau=0,{\bf x}=0,a)
    \eqqcolon
    \rho_{\rm ave}^a,
    \\
    \kappa_{\rm ave}^{(*)}
    (\xi:=(\tau,{\bf x},a),\xi':=(\tau',{\bf x}',a'))
    =&
    \kappa_{\rm ave}^{(*)}
    (\tau-\tau',{\bf x}-{\bf x}',a,0,{\bf 0},a')
    \eqqcolon
    \kappa_{\rm ave}^{aa'(*)}
    (\tau-\tau',{\bf x}-{\bf x}').
\end{align*}
Since the spin-saturated system under consideration 
has the time-reversal symmetry, we have
\begin{align*}
	\mathcal{V}^{\kappa^{*}}_{{\rm KS},aa'}(\tau,{\bf x})
    =
	\mathcal{V}^{\kappa}_{{\rm KS},a'a}(\tau,{\bf x})^{*}.
\end{align*}
Furthermore, the following relation should be satisfied because of the spin symmetry:
\begin{align*}
    \rho_{\rm ave}^\uparrow
    =&
	\rho_{\rm ave}^\downarrow
	\eqqcolon
	\rho_{\rm ave}/2,
	\\
	\mathcal{V}^\rho_{{\rm KS},\uparrow}
    =&
	\mathcal{V}^\rho_{{\rm KS},\downarrow}
    \eqqcolon
	\mathcal{V}^\rho_{{\rm KS}},
    \\
    \kappa_{\rm ave}^{\uparrow\downarrow}(\tau,{\bf x})
    =&
    -\kappa_{\rm ave}^{\downarrow\uparrow}(\tau,{\bf x})
    \eqqcolon
    \kappa_{\rm ave}^{\rm s}(\tau,{\bf x}),
    \\
	\mathcal{V}^\kappa_{{\rm KS},\uparrow\downarrow}(\tau,{\bf x})
    =&
	-\mathcal{V}^\kappa_{{\rm KS},\downarrow\uparrow}(\tau,{\bf x})
    \eqqcolon
	\mathcal{V}^{\kappa,{\rm s}}_{{\rm KS}}(\tau,{\bf x}).
\end{align*}
Here, 
$\rho_{\rm ave}
=\rho_{\rm ave}^\uparrow
+\rho_{\rm ave}^\downarrow$
is the total particle-number density.
Other components for
$\kappa_{\rm ave}^{ab}(\tau,{\bf x})$
and
$\mathcal{V}^\kappa_{{\rm KS},ab}(\tau,{\bf x})$
are assumed to be zero
since we focus on the systems with the spin-singlet condensate.

For convenience for the description of the homogeneous system, 
let us move to the momentum representation.
Then  Eq.~\eqref{eq:jcond_final} 
is reduced to the following two equations for the particle and pairing densities, respectively,
\begin{align}
	\label{eq:jstkappa}
	\begin{cases}
	\displaystyle
    \mathcal{V}_{{\rm KS}}^\rho
    =
    \frac{1}{2}
    \tilde{U}({\bf 0})\rho_{\rm ave}
	\\
	\displaystyle
	\tilde{\mathcal{V}}^{\kappa,{\rm s}}_{{\rm KS}}(P)
    =
    \frac{1}{2}
    \int_{Q}
    \tilde{U}({\bf q}-{\bf p})
    \tilde{\kappa}_{\rm ave}^{\rm s}(Q)^{*}
	\end{cases}.
\end{align}
Here, we have introduced the notations as
$P=(p_{0},{\bf p})$ and $Q=(q_{0},{\bf q})$
with the Matsubara frequencies $p_{0}$ and $q_{0}$, the spacial momenta ${\bf p}$ and ${\bf q}$, and
 $\tilde{\mathcal{V}}^{\kappa,{\rm s}}_{{\rm KS}}(P)$,
$\tilde{U}({\bf p})$ and $\tilde{\kappa}_{\rm ave}^{\rm s}(P)$
are the Fourier transforms of
$\mathcal{V}^{\kappa,{\rm s}}_{{\rm KS}}(\tau,{\bf x})$,
$U({\bf x})$,
and $\kappa_{\rm ave}^{\rm s}(\tau,{\bf x})$, respectively. The following abbreviation is also 
used for the $Q$-`integrations':
\[
\int_Q=\beta^{-1} \sum_{q_{0}}\int d{\bf q}/(2\pi)^3.
\] 
Notice that the right-hand side of Eq.~\eqref{eq:jstkappa} 
and hence $\tilde{\mathcal{V}}^{\kappa,{\rm s}}_{{\rm KS}}(P)$ in the left-hand side are independent of $p_0$.
Thus we write
as $\tilde{\mathcal{V}}^{\kappa,{\rm s}}_{{\rm KS}}(P)=
\tilde{\mathcal{V}}^{\kappa,{\rm s}}_{{\rm KS}}({\bf p})$ from now on.

Now we calculate the equilibrium densities in the
KS system
via the variational equation \eqref{eq:ks_vari} with the obtained 
KS potential Eq.~\eqref{eq:jstkappa}.
However, we have already seen that the solution $\vec{\rho}_{\rm ave}$ 
of Eq.~\eqref{eq:ks_vari} satisfies the self-consistent equation Eq.~\eqref{eq:rhoave_ks}, 
which is now simplified to
\begin{align*}
	\rho_{{\rm ave},i}(\overline{\xi}^i)
	=
	\frac{1}{Z_{\lambda=0}[\vec{\mu}-\vec{\mathcal{V}}_{\rm KS}]}
    \int \mathcal{D}\psi \mathcal{D}\psi^*
    \hat{\rho}_i(\overline{\xi}^i)
    e^{-S_{\lambda=0}[\psi^*, \psi]
    +
    \sum_{j=\rho,\kappa,\kappa^*}
    \int_{\overline{\xi}_{1}^{j}} 
    \left(
    \mu^{j}-\mathcal{V}^{j}_{\rm KS}(\overline{\xi}_{1}^{j})
    \right)
    \hat{\rho}_{j}(\overline{\xi}_{1}^{j})}.
\end{align*}
Furthermore, since the exponent in the integral becomes a bilinear form when Eq.~\eqref{eq:jstkappa} is inserted,
we can perform the path integral analytically to give
\begin{align}
	\label{eq:rhoresult}
    \rho_{\rm ave}
    =&
    \int_{\bf p}
    \left(
    1-\frac{\varepsilon({\bf p})}{E({\bf p})}
    \tanh \frac{\beta E({\bf p})}{2}
    \right),
    \\
    \label{eq:kapparesult}
    \tilde{\kappa}_{\rm ave}^{\rm s}(P)
    =&
    -
    \frac{2\tilde{\mathcal{V}}^{\kappa,{\rm s}}_{{\rm KS}}({\bf p})^*}
    {p_0^2 + E({\bf p})^2},
\end{align}
where $\int_{\bf p}=\int d{\bf p}/{(2\pi)^3}$, 
$
	\varepsilon({\bf p})
	=
	{\bf p}^2/{2}+\mathcal{V}_{{\rm KS}}^\rho-\mu
$, 
and 
$
	E({\bf p})=
	\sqrt{\varepsilon ({\bf p})^2
	+
	|2\tilde{\mathcal{V}}^{\kappa,{\rm s}}_{{\rm KS}}({\bf p})|^2}
$,
as is detailed in Appendix.
Substitution of Eq.~\eqref{eq:kapparesult} into
the second equation of
Eq.~\eqref{eq:jstkappa} in turn leads to
\begin{align}
	\label{eq:gapeq}
	\begin{cases}
	\displaystyle
    \tilde{\mathcal{V}}^{\kappa,{\rm s}}_{{\rm KS}}({\bf p})
    =
    -
    \int_{\bf q}
    \tilde{U}({\bf q}-{\bf p})
    \frac{\tilde{\mathcal{V}}^{\kappa,{\rm s}}_{{\rm KS}}(\bf q)}
    {2\sqrt{\varepsilon ({\bf q})^2+|2\tilde{\mathcal{V}}^{\kappa,{\rm s}}_{{\rm KS}}({\bf q})|^2}}
	\tanh\frac{\beta \sqrt{\varepsilon ({\bf q})^2+|2\tilde{\mathcal{V}}^{\kappa,{\rm s}}_{{\rm KS}}({\bf q})|^2}}{2}
	\\
	\displaystyle
	\varepsilon({\bf p})
	=
	\frac{{\bf p}^2}{2}
	+
	\frac{1}{2}
	\tilde{U}({\bf 0})\rho_{\rm ave}
	-
	\mu
	\end{cases}.
\end{align}
Equation \eqref{eq:kapparesult} implies 
that 
$\tilde{\mathcal{V}}^{\kappa,{\rm s}}_{{\rm KS}}({\bf p})$
is identified with
the energy gap $\Delta^{\rm s}({\bf p})$ in the BCS theory
as $\Delta^{\rm s}({\bf p})=-2    \tilde{\mathcal{V}}^{\kappa,{\rm s}}_{{\rm KS}}({\bf p})^*$. 
With this identification, we find that 
Eq.~\eqref{eq:gapeq} is nothing else than the
celebrated gap equation 
in the BCS theory~\cite{BCS57}.

As is also shown in the Appendix, the Helmholtz free energy is
calculated from Eq.~\eqref{eq:fave_lo2} to yield
\begin{align}
	\label{eq:fave_homo}
	\frac{F_{\rm H}[\vec{\rho}_{\rm ave}]}{V}
	=&
	\frac{\left.\Gamma_{\lambda=0}\right|_{\mathcal{V}=0}[\vec{\rho}_{\rm ave}]}{\beta V}
	+
	\frac{1}{2}
	\tilde{U}(\bm{0})
	\rho_{\mathrm{ave}}^2
	\notag \\
	&-
	\frac{1}{4}
	\int_{\bm{p}, \bm{q}}
	\tilde{U}(\bm{p}-\bm{q})
	\left(
	1-\frac{\varepsilon(\bm{p})}{E(\bm{p})} \tanh \frac{\beta E(\bm{p})}{2}
	\right)
	\left(
	1-\frac{\varepsilon(\bm{q})}{E(\bm{q})} \tanh \frac{\beta E(\bm{q})}{2}
	\right)
	\notag \\
	&+
	\int_{\bm{p}, \bm{q}}
	\tilde{U}(\bm{p}-\bm{q})
	\frac{\tilde{\mathcal{V}}_\mathrm{KS}^{\kappa, s}(\bm{p}) \tilde{\mathcal{V}}_\mathrm{KS}^{\kappa, s}(\bm{q})^*}
	{E(\bm{p})E(\bm{q})} 
	\tanh \frac{\beta E(\bm{p})}{2}
	\tanh \frac{\beta E(\bm{q})}{2}.
\end{align}
where $V=\int d{\bf x}$ is the volume of the system.

Taking the zero temperature limit $\beta \to \infty$ in Eq.~\eqref{eq:fave_homo}, 
the ground-state energy is obtained as
\begin{align}
    \frac{E_{\rm gs}}{V}
	=&
	\frac{T[\vec{\rho}_{\rm gs}]}{V}
	+
	\frac{1}{2}
	\rho^2_{\rm gs}
	\tilde{U}({\bf 0})
	-
	\frac{1}{4}
	\int_{{\bf p},{\bf q}}
	\tilde{U}({\bf p}-{\bf q})
	\left(
	1-\frac{\varepsilon({\bf p})}{E({\bf p})}
	\right)
	\left(
	1-\frac{\varepsilon({\bf q})}{E({\bf q})}
	\right)
	\notag
	\\
	&+
	\frac{1}{4}
	\int_{{\bf p},{\bf q}}
	\tilde{U}({\bf p}-{\bf q})
	\frac{\Delta^{\rm s}({\bf p})\Delta^{\rm s}({\bf q})^*}
	{E({\bf p})E({\bf q})},
\end{align}
Our ground-state energy is identical to that obtained in the BCS theory 
with the identification $\Delta^{\rm s}({\bf p})=-2 \tilde{\mathcal{V}}^{\kappa,{\rm s}}_{{\rm KS}}({\bf p})^*$~\cite{BCS57,sch64,rin80,tin04}. 

\section{Conclusion \label{sec:conclusion}}

On the basis of the effective action formalism, we have developed 
a generalized density-functional theory (DFT) for superfluid systems at finite temperature
in a rigorous way.
The rigorous formulation is combined with the renormalization group method
by introducing a scale parameter that is multiplied to the inter-particle potential,
which is reminiscent of the adiabatic connection method.
A possible difficulty that the spontaneous symmetry breaking (SSB) is 
not described when symmetry-breaking term is absent is nicely circumvented by a suitable
choice of the external fields 
by making use of the functional renormalization group (FRG). 

Then we have established the Hohenberg--Kohn theorem 
with the SSB being taken into account.
By introducing flow-parameter-dependent source terms appropriately fixing the ground-state 
densities during the flow,
we have derived the flow equation 
for the effective action 
for the particle-number and nonlocal pairing densities
where the flow parameter $\lambda$ runs from $0$ to $1$, corresponding to the 
non-interacting and interacting systems, respectively. 

Integrating this flow equation and using the variational equation 
for the
thermal equilibrium densities, we
have arrived at the exact 
self-consistent equation 
to determine the 
Kohn--Sham potential $\mathcal{V}_{\rm KS}^i[\vec{\rho}]$ 
generalized to including the pairing potentials at finite temperature
without introduction of single-particle orbitals.
The resultant Kohn--Sham potential has a nice feature  that it expresses the 
microscopic formulae 
of the external, Hartree, pairing, and exchange-correlation terms, separately.
In particular,  the exchange-correlation term $\mathcal{V}_{\rm KS, xc}^i[\vec{\rho}]$
is  given explicitly in terms of the corresponding density-density correlation functions 
$G_{\rm xc,\lambda}^{(2)}[\vec{\rho}]$.

As a demonstration of the validity of our formulation 
as a microscopic theory of DFT for superfluidity,
we have shown that our 
self-consistent equation with
Kohn--Sham potential leads to the ground-state energy 
of the Hartree--Fock--Bogoliubov theory 
when the correlations are neglected.
And the self-consistent equation is further reduced for short-range
interaction and in the weak-coupling limit to reproduce
the gap equation of the BCS theory
for the spin-singlet pairing.

Although we have shown the validity of our microscopic formulation of DFT based on
FRG by applying it to simple cases,
an advantage of our exact  FRG-DFT formalism lies in the fact 
that it contains the equation~\eqref{eq:g2correlation} with which
the correlation part can be improved systematically.
One of the most effective systematic schemes 
used in the FRG-DFT
is the vertex expansion,
where the Taylor expansion around 
the equilibrium densities are applied to 
the flow equation for the effective action.
In fact, it is shown~\cite{yok18b} 
for  the system without superfluidity that the vertex expansion only
up to the second order gives 
an approximate scheme superior to the random phase approximation (RPA).
Therefore it is naturally expected that a simple vertex expansion applied
to the basic equation in the present work would lead to 
an approximation scheme that incorporates correlations beyond those given by
the quasi-particle RPA.
We hope to report on such an attempt near future.

Our formulation was given to general cases,
i.e. any spatial profiles
and the dependence of the interaction
on the internal degrees of freedom are not imposed.
Such a general formulation may be helpful when applying 
to the superfluidity in various systems. 
Introduction of the nonlocal particle density is an interesting extension of the present formalism, 
with which we can describe the particle-hole and particle-particle correlations on the same footing.
Furthermore, our general formulation for nonlocal pairing
should provide us with a sound basis for
describing pairing phenomena with any symmetry such as
 spin-triplet superfluidity and/or the paring with spin-orbit coupling incorporated.
We hope that there will be a chance to report on such investigations.

\section*{Acknowledgements}
T.~Y.~was supported by the Grants-in-Aid 
for JSPS fellows (Grant No. 20J00644).
This work is supported in part by the Grants-in-Aid for Scientific Research
from JSPS (Nos. 
JP19K03872 
and 
JP19K03824), 
the Yukawa International Program for Quark-hadron Sciences (YIPQS), 
and by 
JSPS-NSFC Bilateral Program for Joint Research Project on 
``Nuclear mass and life for unraveling mysteries of the r-process.''

\appendix

\section{Equilibrium densities and Helmholtz energy of homogeneous systems}

In this appendix we shall derive the equilibrium densities and Helmholtz 
free energy of homogeneous systems.

\subsection{Free propagators}

As was done in Sec.~\ref{sec:homo}, 
we consider a system with the translational, 
time-reversal, and spin symmetries.
The generating functional 
for non-interacting fermions in the presence 
of $\tau$-independent external sources coupled 
to the total particle-number density and spin-singlet pairing density reads:
\begin{equation}
	Z_0 [\vec{J}] 
	=
	\int \mathcal{D}\psi\mathcal{D}\psi^* 
	\e^{
	-S_0[\psi, \psi^*] 
	+J_{\rho} \int_X \hat{\rho}(X) 
	+\int_{X, X'} 2J_{\kappa^s}(\bm{x}-\bm{x}')\delta(\tau-\tau') \hat{\kappa^s}(X,X') 
	+\int_{X, X'} 2J_{\kappa^s}(\bm{x}-\bm{x}')^*\delta(\tau-\tau') \hat{\kappa^s}^{*}(X,X')},
\end{equation}
where $X = (\tau, \bm{x})$, $\int_X \coloneqq \int_0^\beta d\tau \int d\bm{x}$, $\vec{J}=(J_\rho, J_{\kappa^s}(\bm{x}), J_{\kappa^s}(\bm{x})^*)$, 
and
\begin{align}
	\label{eq:s0_app}
	&S_0[\psi, \psi^*]
	=\sum_{s=\uparrow\downarrow} \int_X 
	\psi_s^*(X+\epsilon_\tau) \left(\partial_\tau-\frac{\Delta}{2}\right) \psi_s(X),
	\\
	&\hat{\rho}(X)
	=\sum_{s=\uparrow\downarrow} \psi^*_s(X+\epsilon_\tau)\psi_s(X),
	\\
	&\hat{\kappa^s}(X,X')
	=\frac{1}{2} (\psi_\uparrow(X)\psi_\downarrow(X') -\psi_\downarrow(X)\psi_\uparrow(X') ), 
	\\
	&\hat{\kappa^s}^{*}(X,X')
	=\frac{1}{2} (\psi_\downarrow^*(X')\psi_\uparrow^*(X) -\psi_\uparrow^*(X')\psi_\downarrow^*(X) ).
\end{align}
Here, $X+\epsilon_\tau = (\tau+\epsilon, \bm{x})$, and $\epsilon$ ($>0$) is the time 
step size of the path integral, which will 
 be taken to be zero in the end of the calculation.
We first note 
that the time derivative of 
the Grassmann field $\partial_\tau \psi_s(X)$ in Eq.~\eqref{eq:s0_app} is 
defined as
\begin{equation}
	\partial_\tau \psi_s(X) \coloneqq \frac{\psi_s(X+\epsilon_\tau)-\psi_s(X)}{\epsilon}.
\end{equation}
By the Fourier transformation
\begin{align*}
	\psi_s(X)
	=
	\int_P \e^{\ii P\cdot X} \tilde{\psi}_s(P) 
	=
	\int_P \e^{-\ii\omega_n \tau+\ii \bm{p}\cdot \bm{x}} \tilde{\psi}_s(P), 
	\quad 
	\psi_s^*(X)
	=
	\int_P \e^{-\ii P\cdot X} \tilde{\psi}_s^*(P), 
	\quad 
	J_{\kappa^s}(\bm{x})
	=
	\int_{\bm{p}} \e^{\ii \bm{p}\cdot \bm{x}} \tilde{J}_{\kappa^s}(\bm{p}),
\end{align*}
where $P=(\omega_n, \bm{p})$ with the Matsubara frequency 
$\omega_n=(2n+1)/\beta$, $\int_P \coloneqq \beta^{-1}\sum_{\omega_n} \int d\bm{p}/(2\pi)^{3}$, 
and $\int_{\bm{p}} \coloneqq \int d\bm{p}/(2\pi)^{3}$, 
the generating functional is rewritten in the momentum representation as:
\begin{align}
	\label{eq:gf_mr}
	Z_0[\vec{J}] 
	&=\int \mathcal{D}\tilde{\psi}\mathcal{D}\tilde{\psi}^* 
	\mathrm{exp} 
	\Bigg\{
	\sum_s \int_P 
	\tilde{\psi}_s^*(P)
	\left[
	\frac{\e^{\ii \omega_n \epsilon}-1}{\epsilon} 
	-
	\e^{\ii\omega_n\epsilon} 
	\left(\frac{\bm{p}^2}{2} -J_{\rho}\right) 
	\right]
	\tilde{\psi}_s(P) \notag \\
	&\hspace{5mm}
	+
	\int_P 
	\tilde{J}_{\kappa^s}(\bm{p}) 
	\left[
	\tilde{\psi}_\uparrow(-P)\tilde{\psi}_\downarrow(P) 
	-\tilde{\psi}_\downarrow(-P)\tilde{\psi}_\uparrow(P) 
	\right] 
	+
	\int_P 
	\tilde{J}_{\kappa^s}(\bm{p})^*  
	\left[
	\tilde{\psi}_\downarrow^*(P)\tilde{\psi}_\uparrow^*(-P) 
	-\tilde{\psi}_\uparrow^*(P)\tilde{\psi}_\downarrow^*(-P) 
	\right] 
	\Bigg\}.
\end{align}
The usual treatment of the exponent in the first term may be to make a replacement
 $(\e^{\ii \omega_n \epsilon}-1)/\epsilon \to \ii \omega_n$ 
while leaving $\e^{\ii\omega_n\epsilon} (\bm{p}^2/2 -J_{\rho})$. 
Indeed, nothing is wrong with this practical treatment and
 it will give the correct result.
Instead here we leave all $\epsilon$s until just before 
the end of the calculation for an instructive purpose.

Introducing $\tilde{\Psi}(P) \coloneqq \begin{bmatrix} \tilde{\psi}_\uparrow(P)& \tilde{\psi}_\downarrow^*(-P)\end{bmatrix}^\mathsf{T}$, the exponent of the right-hand side of Eq.~\eqref{eq:gf_mr} is organized in a matrix form 
as in the Nambu--Gor'kov formalism:
\begin{equation}
	Z_0[\vec{J}] =\int \mathcal{D}\tilde{\psi}\mathcal{D}\tilde{\psi}^* 
	\mathrm{exp} \left\{- \int_P \tilde{\Psi}^\dagger(P) 
	\begin{bmatrix} \frac{1-\e^{\ii \omega_n \epsilon}}{\epsilon} + \varepsilon(\bm{p}; J_\rho) \e^{\ii\omega_n\epsilon} &2\tilde{J}_{\kappa^s}(\bm{p})^* \\ 2\tilde{J}_{\kappa^s}(\bm{p}) & -\frac{1-\e^{-\ii \omega_n \epsilon}}{\epsilon} -\varepsilon(\bm{p}; J_\rho) \e^{-\ii\omega_n\epsilon} \end{bmatrix} 
	\tilde{\Psi}(P) \right\}, \label{eq:gf_mr2}
\end{equation}
where $\varepsilon(\bm{p}; J_\rho) \coloneqq \bm{p}^2/2-J_{\rho}$.
By adding external source terms $\sum_{a=1,2} \left[\int_P \tilde{\eta}_a^*(P) \tilde{\Psi}_a(P) + \int_P \tilde{\Psi}_a^*(P) \tilde{\eta}_a(P)\right]$ to the exponent of the right-hand side of Eq.~\eqref{eq:gf_mr2} with $\tilde{\eta}$ and $\tilde{\eta}^*$ being Grassmann variables, and differentiating the generating functional with respect to $\tilde{\eta}$ and $\tilde{\eta}^*$, the propagators are obtained in a matrix form:
\begin{align}
	&\Braket{\begin{bmatrix} 
		\tilde{\psi}_\uparrow(P) \tilde{\psi}_\uparrow^*(P) & 
		\tilde{\psi}_\uparrow(P)\tilde{\psi}_\downarrow(-P) \\ 
		\tilde{\psi}_\downarrow^*(-P) \tilde{\psi}_\uparrow^*(P) & 
		\tilde{\psi}_\downarrow^*(-P) \tilde{\psi}_\downarrow(-P)
	\end{bmatrix} }_{\vec{J}} 
	= 
	\begin{bmatrix} 
		\frac{1-\e^{\ii \omega_n \epsilon}}{\epsilon} + \varepsilon(\bm{p};J_\rho) \e^{\ii\omega_n\epsilon} &
		2\tilde{J}_{\kappa^s}(\bm{p})^* \\ 
		2\tilde{J}_{\kappa^s}(\bm{p}) & 
		-\frac{1-\e^{-\ii \omega_n \epsilon}}{\epsilon} -\varepsilon(\bm{p};J_\rho) \e^{-\ii\omega_n\epsilon} 
	\end{bmatrix}^{-1} 
	\notag \\
	&=\frac{1}{
	\left(
	\frac{2\sqrt{1-\epsilon\varepsilon(\bm{p};J_\rho)}}{\epsilon}
	\sin\frac{\omega_n\epsilon}{2}
	\right)^2
	+
	\left.E(\bm{p};\vec{J})\right.^2} 
	\begin{bmatrix} 
		\frac{2\ii\e^{-\ii \frac{\omega_n\epsilon}{2}}}{\epsilon}\sin\frac{\omega_n\epsilon}{2} +\varepsilon(\bm{p};J_\rho) \e^{-\ii\omega_n\epsilon} &
		2\tilde{J}_{\kappa^s}(\bm{p})^* \\ 
		2\tilde{J}_{\kappa^s}(\bm{p}) &
		\frac{2\ii\e^{\ii \frac{\omega_n\epsilon}{2}}}{\epsilon}\sin\frac{\omega_n\epsilon}{2}  - \varepsilon(\bm{p};J_\rho) \e^{\ii\omega_n\epsilon} 
	\end{bmatrix},
\end{align}
where $E(\bm{p};\vec{J}) \coloneqq \sqrt{\varepsilon(\bm{p};J_\rho)^2 + \left|2\tilde{J}_{\kappa^s}(\bm{p})\right|^2}$, and $\braket{\mathcal{O}[\psi,\psi^*]}_{\vec{J}}$ is defined as
\begin{align}
	&\braket{\mathcal{O}[\psi,\psi^*]}_{\vec{J}} 
	\notag \\
	&
	\coloneqq \frac{1}{Z_0[\vec{J}]} 
	\int \mathcal{D}\psi\mathcal{D}\psi^* \mathcal{O}[\psi,\psi^*] 
	\e^{-S_0[\psi, \psi^*] +J_{\rho} \int_X \hat{\rho}(X) + \int_{X,X'} 2J_{\kappa^s}(\bm{x}-\bm{x}') \delta(\tau-\tau') \hat{\kappa^s}(X,X') +\int_{X X'} 2J_{\kappa}(\bm{x}-\bm{x}')^* \delta(\tau-\tau') \hat{\kappa^s}^*(X,X') }.
\end{align}
with an arbitrary functional
$\mathcal{O}[\psi, \psi^*]$ of $\psi$ and $\psi^*$ including the Fourier transforms of them.

\subsection{Equilibrium densities}
From Eq.~\eqref{eq:rhoave_ks} the equilibrium densities satisfy
\begin{align}
	\rho_{\rm ave}
	&=
	\braket{
	\hat{\rho}(X)
	}_{\vec{\mu}-\vec{\mathcal{V}}_{\mathrm{KS}}},
	\\
	\kappa_{\rm ave}^{s (*)}(X-X')
	&=
	\braket{
	\hat{\kappa^s}^{(*)}(X,X')
	}_{\vec{\mu}-\vec{\mathcal{V}}_{\mathrm{KS}}}.
\end{align}
The equilibrium total particle-number density is calculated as follows:
\begin{align}
	\label{eq:rhoave_app}
	\rho_\mathrm{ave}
	&=
	\sum_s
	\braket{\psi_s^*(X+\epsilon_\tau)\psi_s(X)}_{\vec{\mu}-\vec{\mathcal{V}}_{\mathrm{KS}}} 
	\notag \\
	&=
	\int_P \e^{\ii \omega_n\epsilon} \sum_s 
	\braket{\tilde{\psi}_s^*(P)\tilde{\psi}_s(P)}_{\vec{\mu}-\vec{\mathcal{V}}_{\mathrm{KS}}} 
	\notag \\
	&=
	2\int_P 
	\frac{-\frac{2\ii\e^{\ii \frac{\omega_n \epsilon}{2}}}{\epsilon}\sin\frac{\omega_n\epsilon}{2} -\varepsilon(\bm{p}) }
	{\left(\frac{2\sqrt{1-\epsilon\varepsilon(\bm{p})}}{\epsilon}\sin\frac{\omega_n\epsilon}{2}\right)^2+E(\bm{p})^2}, 
\end{align}
where 
$\varepsilon(\bm{p})\coloneqq \varepsilon(\bm{p};\mu-\mathcal{V}_\mathrm{KS}^\rho) = \bm{p}^2/2+\mathcal{V}_\mathrm{KS}^\rho-\mu$, 
and $E(\bm{p}) \coloneqq E(\bm{p}; \vec{\mu}-\vec{\mathcal{V}}_\mathrm{KS})=\sqrt{\varepsilon(\bm{p})^2+\left|2\tilde{\mathcal{V}}_\mathrm{KS}^{\kappa, s}(\bm{p})\right|^2}$. To calculate the Matsubara frequency summation in Eq.~\eqref{eq:rhoave_app}, let us introduce a complex function
\begin{align*}
	h(z) 
	= 
	\frac{-\frac{2\ii\e^{\ii \frac{z \epsilon}{2}}}{\epsilon}\sin\frac{z\epsilon}{2} -\varepsilon }
	{\left(\frac{2\sqrt{1-\epsilon\varepsilon}}{\epsilon}\sin\frac{z\epsilon}{2}\right)^2+E^2} 
\end{align*}
and a weighting function 
\begin{align*}
	g(z)
	=
	-\beta n_F(z) 
	= 
	\frac{-\beta}{1+\e^{\beta z}}.
\end{align*}
It is observed that the product $h(-iz)g(z)$ decays sufficiently fast as $|z|\rightarrow \infty$, and $h(-iz)$ has simple poles at $z=\pm \frac{2}{\epsilon} \sinh^{-1} \frac{\epsilon E}{2\sqrt{1-\epsilon\varepsilon}}$. Therefore,
\begin{align}
	\frac{1}{\beta}\sum_{\omega_n} h(\omega_n) 
	&=
	-\frac{1}{\beta} 
	\mathrm{Res}
	\left(
	h(-iz)g(z), 
	z=\frac{2}{\epsilon} \sinh^{-1} \frac{\epsilon E}{2\sqrt{1-\epsilon\varepsilon}}
	\right) 
	\notag \\
	&\hspace{1cm} 
	-\frac{1}{\beta} 
	\mathrm{Res}
	\left(
	h(-iz)g(z), 
	z=-\frac{2}{\epsilon} \sinh^{-1} \frac{\epsilon E}{2\sqrt{1-\epsilon\varepsilon}}
	\right) 
	\notag \\
	&
	=
	\frac{E+\varepsilon}{2E} n_F(E) 
	+ 
	\frac{E-\varepsilon}{2E} n_F(-E) 
	\notag \\
	&=
	\frac{1}{2}
	\left(
	1-\frac{\varepsilon}{E}\tanh \frac{\beta E}{2} 
	\right).
\end{align}
In the second to last line, we have taken the limit $\epsilon \to 0$ implicitly. Consequently, we arrive at
\begin{equation}
	\rho_{\mathrm{ave}} 
	= 
	\int_{\bm{p}} 
	\left( 
	1-\frac{\varepsilon(\bm{p})}{E(\bm{p})} \tanh \frac{\beta E(\bm{p})}{2}
	\right).
\end{equation}
In the same way, the equilibrium spin-singlet pairing density is calculated as follows:
\begin{align}
	\kappa_{\mathrm{ave}}^s(X-X') 
	&= 
	\frac{1}{2} 
	\braket{
	\psi_\uparrow(X)\psi_\downarrow(X') -\psi_\downarrow(X)\psi_\uparrow(X')
	}_{\vec{\mu}-\vec{\mathcal{V}}_{\mathrm{KS}}} 
	\notag \\
	&= 
	\frac{1}{2} 
	\int_P 
	\e^{\ii P \cdot(X-X')}
	\left[
	\braket{
	\tilde{\psi}_\uparrow(P)\tilde{\psi}_\downarrow(-P)
	}_{\vec{\mu}-\vec{\mathcal{V}}_{\mathrm{KS}}}
	-
	\braket{
	\tilde{\psi}_\downarrow(P)\tilde{\psi}_\uparrow(-P)
	}_{\vec{\mu}-\vec{\mathcal{V}}_{\mathrm{KS}}}
	\right] 
	\notag \\
	&= 
	\int_{P} 
	\e^{\ii P \cdot(X-X')}
	\frac{-2\tilde{\mathcal{V}}_\mathrm{KS}^{\kappa, s}(\bm{p})^*}
	{\left(\frac{2\sqrt{1-\epsilon\varepsilon(\bm{p})}}{\epsilon}\sin\frac{\omega_n\epsilon}{2}\right)^2+E(\bm{p})^2} 
	\notag \\
	& 
	=
	\int_{\bm{p}} 
	\e^{\ii \bm{p}\cdot(\bm{x}-\bm{x}')} 
	\frac{-\tilde{\mathcal{V}}_\mathrm{KS}^{\kappa, s}(\bm{p})^*}{E(\bm{p})}
	\left(
	\frac{\e^{-E(\bm{p}) \left|\tau-\tau'\right|}}{1+\e^{-\beta E(\bm{p})}}
	-
	\frac{\e^{E(\bm{p}) \left|\tau-\tau'\right|}}{1+\e^{\beta E(\bm{p})}}
	\right).
\end{align}
In particular, the local spin-singlet pairing density in equilibrium is given by
\begin{align}
	\kappa_{\rm ave}^s(0) 
	= 
	\int_{\bm{p}} 
	\frac{-\tilde{\mathcal{V}}_\mathrm{KS}^{\kappa, s}(\bm{p})^*}{E(\bm{p})}
	\tanh \frac{\beta E(\bm{p})}{2}.
\end{align}

\subsection{Helmholtz free energy}
From Eq.~\eqref{eq:fave_lo2} the Helmholtz energy in the lowest order approximation is given by
\begin{align}
	\label{eq:fave_app}
	F_{\rm H}[\vec{\rho}_{\rm ave}]
	=&
	\frac{\Gamma_{\lambda=0}[\vec{\rho}_{\rm ave}]}{\beta}
	+
	\frac{1}{2\beta}
	\int_{\overline{\xi}^\rho_1,\overline{\xi}^\rho_2}
	\mathcal{U}_{\lambda=1}(\overline{\xi}^\rho_1,\overline{\xi}^\rho_2)
	\rho_{\rm ave}(\overline{\xi}^\rho_1)\rho_{\rm ave}(\overline{\xi}^\rho_2)
	\notag \\
	&
	-
	\frac{1}{2\beta}
	\int_{\overline{\xi}^\rho_1,\overline{\xi}^\rho_2}
	\mathcal{U}_{\lambda=1}(\overline{\xi}^\rho_1,\overline{\xi}^\rho_2)
	\braket{
	\psi^*(\xi_1+\epsilon_\tau+\epsilon'_\tau)
	\psi(\xi_2)
	}_{\vec{\mu}-\vec{\mathcal{V}}_{\mathrm{KS}}}
	\braket{
	\psi^*(\xi_2+\epsilon_\tau+\epsilon'_\tau)
	\psi(\xi_1)
	}_{\vec{\mu}-\vec{\mathcal{V}}_{\mathrm{KS}}}
	\notag
	\\
	&
	+
	\frac{1}{2\beta}
	\int_{\overline{\xi}^\rho_1,\overline{\xi}^\rho_2}
	\mathcal{U}_{\lambda=1}(\overline{\xi}^\rho_1,\overline{\xi}^\rho_2)
	\kappa^*_{\rm ave}(\overline{\xi}_{12}^{\kappa^*})
	\kappa_{\rm ave}(\overline{\xi}_{12}^{\kappa})
	|_{\overline{\xi}_{12}^{\kappa^{(*)}}=({\overline{\xi}^\rho_1,\overline{\xi}^\rho_2})}.
\end{align}
The second term of the right-hand side of Eq.~\eqref{eq:fave_app} is evaluated as follows:
\begin{align}
	\frac{1}{2\beta}
	\int_{\overline{\xi}^\rho_1, \overline{\xi}^\rho_2} 
	\mathcal{U}_{\lambda=1}(\overline{\xi}^\rho_1, \overline{\xi}^\rho_2) 
	\rho_{\mathrm{ave}}(\overline{\xi}^\rho_1) 
	\rho_{\mathrm{ave}}(\overline{\xi}^\rho_2)
	=
	\frac{1}{2}
	\int_{\bm{x}_1, \bm{x}_2}
	U(\bm{x}_1-\bm{x}_2)
	\sum_{s_1=\uparrow\downarrow}
	\rho_{\mathrm{ave}}^{s_1}
	\sum_{s_2=\uparrow\downarrow}
	\rho_{\mathrm{ave}}^{s_2} 
	=
	\frac{V}{2}
	\tilde{U}(\bm{0})
	\rho_{\mathrm{ave}}^2,
\end{align}
where $V = \int d\bm{x}$ is the volume and $\tilde{U}(\bm{p})=\int d\bm{x} \e^{-\ii \bm{p}\cdot\bm{x}} U(\bm{x})$. The third term is evaluated as follows:
\begin{align}
	&-\frac{1}{2\beta}
	\int_{\overline{\xi}^\rho_1, \overline{\xi}^\rho_2} 
	\mathcal{U}_{\lambda=1}(\overline{\xi}^\rho_1, \overline{\xi}^\rho_2) 
	\langle
	\psi^*(\xi_1+\epsilon_\tau+\epsilon'_\tau)
	\psi(\xi_2)
	\rangle_{\vec{\mu}-\vec{\mathcal{V}}_{\mathrm{KS}}}
	\langle
	\psi^*(\xi_2+\epsilon_\tau+\epsilon'_\tau)
	\psi(\xi_1)
	\rangle_{\vec{\mu}-\vec{\mathcal{V}}_{\mathrm{KS}}}
	\notag \\
	&=
	-\frac{1}{2}
	\int_{\bm{x}_1, \bm{x}_2}
	U(\bm{x}_1-\bm{x}_2)
	\sum_{s_1, s_2=\uparrow\downarrow}
	\left[
	\braket{
	\psi^*_{s_1}(X_1+\epsilon_\tau+\epsilon'_\tau)
	\psi_{s_2}(X_2)
	}_{\vec{\mu}-\vec{\mathcal{V}}_{\mathrm{KS}}}
	\braket{
	\psi^*_{s_2}(\xi_2+\epsilon_\tau+\epsilon'_\tau)
	\psi_{s_1}(\xi_1)
	}_{\vec{\mu}-\vec{\mathcal{V}}_{\mathrm{KS}}}
	\right]_{\tau_1=\tau_2}
	\notag \\
	&=
	-\frac{V}{2}
	\int_{\bm{p}, \bm{q}}
	\tilde{U}(\bm{p}-\bm{q})
	\sum_{s=\uparrow\downarrow}
	\frac{1}{\beta}
	\sum_{m}
	\e^{\ii\omega_n(\epsilon+\epsilon')}
	\braket{
	\tilde{\psi}^*_{s}(\omega_m, \bm{p})
	\tilde{\psi}_{s}(\omega_m, \bm{p})
	}_{\vec{\mu}-\vec{\mathcal{V}}_{\mathrm{KS}}}
	\notag \\
	&\hspace{7cm}
	\times
	\frac{1}{\beta}
	\sum_{n}
	\e^{\ii\omega_n(\epsilon+\epsilon')}
	\braket{
	\tilde{\psi}^*_{s}(\omega_n, \bm{q})
	\tilde{\psi}_{s}(\omega_n, \bm{q})
	}_{\vec{\mu}-\vec{\mathcal{V}}_{\mathrm{KS}}}
	\notag \\
	&=
	-\frac{V}{4}
	\int_{\bm{p}, \bm{q}}
	\tilde{U}(\bm{p}-\bm{q})
	\left(
	1-\frac{\varepsilon(\bm{p})}{E(\bm{p})} \tanh \frac{\beta E(\bm{p})}{2}
	\right)
	\left(
	1-\frac{\varepsilon(\bm{q})}{E(\bm{q})} \tanh \frac{\beta E(\bm{q})}{2}
	\right).
\end{align}
The fourth term is evaluated as follows:
\begin{align}
	&\frac{1}{2\beta}
	\int_{\overline{\xi}^\rho_1,\overline{\xi}^\rho_2}
	\mathcal{U}_{\lambda=1}(\overline{\xi}^\rho_1,\overline{\xi}^\rho_2)
	\kappa^{*}_{\rm ave}(\overline{\xi}_{12}^{\kappa^*})
	\kappa_{\rm ave}(\overline{\xi}_{12}^{\kappa})
	|_{\overline{\xi}_{12}^{\kappa^{(*)}}=({\overline{\xi}^\rho_1,\overline{\xi}^\rho_2})}
	\notag \\
	&=
	\frac{1}{2}
	\int_{\bm{x}_1, \bm{x}_2}
	U(\bm{x}_1-\bm{x}_2)
	\left[
	\kappa_{\rm ave}^{\uparrow\downarrow *}(X_1,X_2)
	\kappa_{\rm ave}^{\uparrow\downarrow}(X_1,X_2)
	+
	\kappa_{\rm ave}^{\downarrow\uparrow *}(X_1,X_2)
	\kappa_{\rm ave}^{\downarrow\uparrow}(X_1,X_2)
	\right]_{\tau_1=\tau_2}
	\notag \\
	&=
	\int_{\bm{x}_1, \bm{x}_2}
	U(\bm{x}_1-\bm{x}_2)
	\left.
	\kappa_{\rm ave}^{s *}(X_1,X_2)
	\right|_{\tau_1=\tau_2}
	\left.
	\kappa_{\rm ave}^{s}(X_1,X_2)
	\right|_{\tau_1=\tau_2}
	\notag \\
	&=
	V
	\int_{\bm{p}, \bm{q}}
	\tilde{U}(\bm{p}-\bm{q})
	\frac{\tilde{\mathcal{V}}_\mathrm{KS}^{\kappa, s}(\bm{p}) \tilde{\mathcal{V}}_\mathrm{KS}^{\kappa, s}(\bm{q})^*}
	{E(\bm{p})E(\bm{q})} 
	\tanh \frac{\beta E(\bm{p})}{2}
	\tanh \frac{\beta E(\bm{q})}{2}.
\end{align}
Consequently, we have
\begin{align}
	\frac{F_{\rm H}[\vec{\rho}_{\rm ave}]}{V}
	=&
	\frac{\left.\Gamma_{\lambda=0}\right|_{\mathcal{V}=0}[\vec{\rho}_{\rm ave}]}{\beta V}
	+
	\frac{1}{2}
	\tilde{U}(\bm{0})
	\rho_{\mathrm{ave}}^2
	\notag \\
	&-
	\frac{1}{4}
	\int_{\bm{p}, \bm{q}}
	\tilde{U}(\bm{p}-\bm{q})
	\left(
	1-\frac{\varepsilon(\bm{p})}{E(\bm{p})} \tanh \frac{\beta E(\bm{p})}{2}
	\right)
	\left(
	1-\frac{\varepsilon(\bm{q})}{E(\bm{q})} \tanh \frac{\beta E(\bm{q})}{2}
	\right)
	\notag \\
	&+
	\int_{\bm{p}, \bm{q}}
	\tilde{U}(\bm{p}-\bm{q})
	\frac{\tilde{\mathcal{V}}_\mathrm{KS}^{\kappa, s}(\bm{p}) \tilde{\mathcal{V}}_\mathrm{KS}^{\kappa, s}(\bm{q})^*}
	{E(\bm{p})E(\bm{q})} 
	\tanh \frac{\beta E(\bm{p})}{2}
	\tanh \frac{\beta E(\bm{q})}{2}.
\end{align}

\bibliographystyle{ptephy}

\bibliography{references}
\end{document}